\documentclass[a4paper,11pt]{article}
\pdfoutput=1 

\usepackage{jheppub} 
\usepackage{mathrsfs,graphicx,rotating,amsmath,amsfonts,mathtools,booktabs,amssymb,wasysym,caption}
\usepackage[fleqn]{nccmath}
\usepackage{hyperref}
\usepackage{slashed}
\usepackage{natbib}
\usepackage[table,xcdraw,dvipsnames]{xcolor}
\usepackage{graphicx}
\usepackage{bbold}
\usepackage[utf8x]{inputenc}
\usepackage[english]{babel}
\usepackage{multirow,multicol}
\usepackage{epstopdf}
\usepackage{bbm}
\usepackage{appendix}
\usepackage{libertine}
\usepackage{braket}
\usepackage{wasysym}
\usepackage{empheq}
\usepackage{cancel}
\usepackage{enumitem}
\usepackage{mathrsfs}
\usepackage{lmodern}
\usepackage{tabularx}
\usepackage{multicol}
\usepackage{color}
\usepackage{mathtools}

\newcommand{\D}{{\rm d}}
\newcommand{\e}{{\rm e}}
\renewcommand\({\left(}
\renewcommand\){\right)}
\renewcommand\[{\left[}
\renewcommand\]{\right]}

\newcommand{\Mpl}{M_{\text{Pl}}}

\title{\boldmath Effective Field Theory for the Perturbations of a Slowly Rotating Black Hole}

\author[a]{Lam Hui,}
\author[a]{Alessandro Podo,}
\author[a,b]{Luca Santoni,}
\author[c,d]{Enrico Trincherini}

\affiliation[a]{Department of Physics, Center for Theoretical Physics, Columbia University, New York, NY 10027, USA}
\affiliation[b]{ICTP, International Centre for Theoretical Physics, Strada Costiera 11, 34151, Trieste, Italy}
\affiliation[c]{Scuola Normale Superiore, Piazza dei Cavalieri 7, 56126, Pisa, Italy}
\affiliation[d]{Istituto Nazionale di Fisica Nucleare (INFN) - Sezione di Pisa \\ Polo Fibonacci Largo B. Pontecorvo, 3, I-56127 Pisa, Italy}

\emailAdd{lh399@columbia.edu}
\emailAdd{ap3964@columbia.edu}
\emailAdd{lsantoni@ictp.it}
\emailAdd{enrico.trincherini@sns.it}

\abstract{We develop the effective theory for perturbations around
  black holes with scalar hair, in two directions. 
First, we show that the scalar-Gauss--Bonnet theory, often used as an
example exhibiting scalar black hole hair, can be deformed by galileon
operators leading to order unity changes to its predictions. The
effective theory for perturbations thus provides an efficient framework for
describing and constraining broad classes of scalar-tensor theories, of which the addition of  galileon operators is  an example. 
Second, we extend the effective theory to perturbations around an
axisymmetric, slowly rotating black hole, at linear order
in the black hole spin. We also discuss the inclusion of
parity-breaking operators in the effective theory.
}

\begin{document}
\maketitle


\newpage

\section{Introduction}

Future gravitational wave experiments are expected to find not only a
large number of black hole  merger events, but also to measure their
gravitational wave signals to high precision. One exciting prospect is
the high quality data on black hole ring-down, sufficient to detect
and measure multiple quasi-normal modes (QNMs) \cite{Baibhav:2019rsa,Giesler:2019uxc,Baibhav:2020tma}.
Each quasi-normal mode is characterized by the real and imaginary
parts of its frequency, and thus one could put to the test the general
relativistic prediction that black holes are completely characterized
by their mass and spin (in addition to charge, which is generally
expected to vanish). There are several different ways how such a test
can be carried out. (1) The simplest one is a null test: are two numbers
(mass and spin) sufficient to describe the observed quasi-normal
spectrum? Beyond that, it is useful to test for the possible presence
of deviations. (2) Concrete models help guide our thinking on the form of
possible deviations and the associated model testing. 
The scalar-Gauss--Bonnet model (and its variations) is a popular
example. It circumvents  earlier no-scalar-hair theorems \cite{Hui:2012qt,Sotiriou:2013qea,Herdeiro:2015waa,Creminelli:2020lxn}
and gives interesting predictions for the background geometry and its
perturbations, see e.g.~\cite{Kanti:1995vq,Sotiriou:2014pfa,Blazquez-Salcedo:2016enn,Antoniou:2017hxj,Witek:2018dmd,Delgado:2020rev}. Another example is scalar-Chern--Simons model, see e.g.~\cite{Smith:2007jm, Grumiller:2007rv, Yunes:2007ss,  Yunes:2009hc,Konno:2009kg, Cardoso:2009pk, Molina:2010fb, Pani:2011gy, Yagi:2012ya, Stein:2013wza, Kimura:2018nxk,Wagle:2021tam,Srivastava:2021imr}. (3) A third approach is to parametrize the deviations from
general relativistic expectations in a model-independent way. 
This can take the form of a phenomenological parametrization such as~\cite{Loutrel:2014vja,Cardoso:2019mqo,McManus:2019ulj,Glampedakis:2019dqh,Silva:2019scu, Maselli:2019mjd}, or a parametrization at the level of the action governing the
dynamics of black hole perturbations \cite{Endlich:2017tqa,Tattersall:2017erk,Franciolini:2018uyq,Cano:2020cao,Cano:2021myl}. It is the last approach that is
the subject of this paper. 
One main reason for our choice is that, as opposed to a
phenomenological parametrization of possible deviations from general
relativity, a parametrization at the level of an effective Lagrangian
makes more transparent which types of deformations correspond to
theories that respect physical principles, such as locality and
diffeomorphism invariance, and makes it easier to connect UV theories
with observations, by a systematic matching procedure.

An effective field theory for perturbations around a spherically symmetric spacetime with non-trivial scalar
background was developed in \cite{Franciolini:2018uyq,Franciolini:2018aad}. To orient ourselves,
it is helpful to think of a perhaps more familiar example: the
effective theory for
perturbations around an inflating universe \cite{Cheung:2007st}. The goal of such a
theory is not to explain where the (inflation) background comes from;
rather, one takes the background as given and proceeds to write down
an action that governs the dynamics of the perturbations, guided by
the notions of symmetries and derivative expansion.
To formulate an effective theory for black hole perturbations, we
proceed in an analogous manner. It is assumed the black hole has a
scalar background with a non-trivial radial profile, much like the
inflaton having a non-trivial temporal profile. Exactly what (UV) physics gives rise to this scalar hair is
immaterial for the construction of the effective theory. The important
point is that one can choose a unitary gauge in which
all the perturbations reside in the metric. 
In this gauge, we use the invariance under $t, \theta,
\phi$-diffeomorphisms as the guiding principle for writing down
the effective (IR) theory.

This strategy was adopted in \cite{Franciolini:2018uyq} to construct 
an effective theory for perturbations around a spherically symmetric
black hole with scalar hair. In this paper, we wish to address two
follow-up questions: (1) What is the utility of an effective theory if
known black hole hair solutions generally make use of a limited set of
interactions (e.g.~scalar-Gauss--Bonnet, or scalar-Chern--Simons)---why
not work directly with the UV theory instead of dealing with an effective IR theory?
(2) Given that black holes in nature have angular momentum,
can the effective theory be generalized to describe rotating black
holes? We answer the first question by showing how the
scalar-Gauss--Bonnet model can be deformed to yield a wide variety of
predictions; an effective theory description is thus a useful way
to parametrize the possibilities. We answer the second question by
showing how the formalism of \cite{Franciolini:2018uyq} can be adapted in a simple way to
slowly rotating black holes. Black holes with a substantial spin is
not as straightforward to work with---recall that even in general
relativity, the only known way to write down separable equations of
perturbations around a Kerr background is to use 
the Newman--Penrose variables, and that there is no known action formulation
for the associated Teukolsky equation.

An outline of the paper is as follows. 
In Section \ref{EFTmotivations}, we go over the motivations for 
the effective theory approach, using the scalar-Gauss--Bonnet model and
its deformations as an example (focusing for simplicity on
non-rotating black holes). In particular, we  show how  in the eikonal
limit a judicious choice of the couplings in the effective  theory
allows one to recover isospectrality, which is usually  broken in
theories beyond general relativity. 
The point is {\it not} to motivate this rather finely-tuned choice of
couplings, but to illustrate how a wide range of behavior is possible
by deforming the scalar-Gauss--Bonnet model.
We then develop the effective theory for
perturbations around slowly rotating black holes
in Section \ref{sec:EFTsrbh}.
We discuss cases with parity-breaking operators in Section
\ref{sec:EFTCS}, and conclude in Section~\ref{sec:conclusions}.
It is worth stressing that the resulting equations from the effective
theory are often difficult to solve. The WKB approach provides an analytic and simple, even if
approximate, way to deduce the quasi-normal spectrum \cite{Schutz:1985km,Iyer:1986np,Seidel:1989bp}. In the even sector, however, one in general obtains a coupled set of equations governing the scalar and tensor modes and the application of the WKB approach is not as straightforward. In a follow-up paper \cite{uspaper2}, we will show how to analytically compute the (approximate) quasi-normal mode spectrum in such a coupled system, in a spirit similar to the WKB approximation.

\paragraph{Conventions:}  We work in mostly-plus signature for the metric, $(-,+,+,+)$, and in units where $c=\hbar=1$. We denote the reduced Planck mass with $\Mpl=(8\pi G)^{-1/2}$, where $G$ is the Newton's constant. We  use Greek indices $\mu, \nu, \ldots$ to denote $4$-dimensional spacetime coordinates, while we use Latin indices $a,b,c,\ldots$ for temporal and angular coordinates $(t,\theta,\phi)$ only.  Latin indices $i,j,\ldots$ label instead the angular coordinates $(\theta,\phi)$ on the $S^2$-sphere.

\section{Motivations for an effective theory}
\label{EFTmotivations}

\subsection{Black holes with galileon hair}

We discuss here an explicit example of hairy black hole where, in addition to the Gauss--Bonnet coupling, the Lagrangian for the scalar  contains another operator describing  a particular higher derivative self-interaction of the field, which we show  can induce order-one corrections to the solution of~\cite{Sotiriou:2014pfa}, where this operator is not included. We will use this example as a  motivation for  introducing an effective theory to study  perturbations around black holes with hair: since the operator considered below is just one particular example of a much larger set of  operators that can in principle affect the linearized dynamics for the perturbations and the observables, a model-independent approach is, in this sense, valuable.

For simplicity,  the example discussed here is in the context of non-rotating black holes, though it can be straightforwardly generalized to the case of slowly rotating spacetimes.

\subsubsection{Some preliminary scaling  considerations}

Let us consider the  action
\begin{equation}
S = \int \D^4 x \sqrt{-g} \left[
\frac{\Mpl^2}{2}R - \frac{1}{2} (\partial\Phi)^2 + g_3\frac{(\partial\Phi)^2\square\Phi}{\Lambda^3}  + \alpha \Mpl \Phi \mathcal{G}
\right] \, ,
\label{theory}
\end{equation}
In the limit $g_3=0$ one recovers the well studied case of \cite{Sotiriou:2014pfa,Blazquez-Salcedo:2016enn}, where the scalar field $\Phi$ is linearly coupled to the Gauss--Bonnet term $\mathcal{G}$,
\begin{equation}
\mathcal{G} \equiv R^{\mu\nu\rho\sigma}R_{\mu\nu\rho\sigma} - 4R^{\mu\nu}R_{\mu\nu} + R^2 \, .
\label{GBdefinition}
\end{equation}
In~\eqref{theory}, we added as an additional operator the cubic galileon self-interaction $(\partial\Phi)^2\square\Phi$ \cite{Nicolis:2008in}, suppressed by a scale $\Lambda$ to be fixed later. A consistent low-energy Effective Field Theory (EFT) will of course contain infinitely many irrelevant operators whose coefficients scale accordingly to naive dimensional analysis. In this section, however, our goal is to show     
that at least some of those operators can be relevant in describing the background solution around black holes and the dynamics of perturbations in addition to the scalar Gauss--Bonnet (sGB) coupling. For this purpose, it is enough to focus on a simple model, though there are in fact other operators that can be considered as well \cite{Noller:2019chl}.

The choice of defining the coefficient of the linear coupling between $\Phi$ and the Gauss--Bonnet invariant in terms of a coupling $\alpha$ with the dimension of length squared, as in~\eqref{theory}, is common in the literature. It is particularly convenient because it immediately identifies the typical length below which corrections from the Gauss--Bonnet coupling become relevant. For a black hole, for example, it is easy to see that the space-time geometry close to the horizon is modified by corrections ${\cal O}(\alpha/r_s^2)$, where $r_s$ defines the position of the black hole horizon. Therefore, we find useful to introduce the dimensionless coupling
\begin{equation}
\tilde{\alpha} \equiv \frac{\alpha}{r_s^2} \, .
\end{equation}
The definition adopted in~\eqref{theory}, on the other hand, makes less transparent another relevant scale associated with that coupling. The sGB operator contains a series of irrelevant interactions between the scalar field and an arbitrary number of gravitons. The leading one is a cubic vertex, schematically of the form $\partial^2 \Phi \, \partial h \, \partial h$. When the graviton field is canonically normalized, the interaction is a dimension 7 operator 
\begin{equation}
\frac{1}{\Lambda_\alpha^3} \partial^2 \Phi \, \partial h_c \, \partial h_c 
\end{equation}
where $\Lambda_\alpha$ is the energy at which the cubic interaction becomes strongly coupled. In terms of $\alpha$ one immediately gets 
\begin{equation}
\frac{1}{\Lambda_\alpha} = \left( \frac{\alpha}{\Mpl} \right)^{1/3} \, .
\end{equation}
From now on, we will assume that the coupling $\alpha$ has the largest value allowed by present observations. The strongest bound on $\alpha$  comes from the absence of any signal of scalar wave dipolar emission, sourced by that coupling, in black hole merger data.
A recent analysis finds $\alpha \lesssim (1.7 \, {\rm km})^2$ \cite{Perkins:2021mhb} (see also Refs.~\cite{Nair:2019iur,Witek:2018dmd}).
In terms of the associated strong coupling scale the bound becomes  
\begin{equation}
\Lambda_\alpha \gtrsim 10^{12} \, {\rm km^{-1}} \, .
\end{equation}
When these values are saturated, the coefficient parametrizing deviations from general relativity predictions for black holes in the LIGO-Virgo band is at most $\tilde{\alpha} \sim 0.1 - 0.01$.

We now turn to estimate the size of the effects induced by the additional galileon interaction. To do so, we have to fix the value of $\Lambda$ in \eqref{theory}, taking for simplicity $g_3 \sim {\cal O}(1)$. We will make a natural assumption: all the leading interactions involving the scalar $\Phi$ become strong at the same scale. This implies that we have to set $\Lambda \sim \Lambda_\alpha$. From now on we will work under this assumption and discuss  the implications of the cubic galileon coupling $g_{3}$ on the hairy black hole solution.

From the scalar's equations of motion, $\square\Phi(1+ \frac{\square\Phi}{\Lambda^3}+ \ldots) \sim \alpha \Mpl \mathcal{G}$, one can see that the presence of the cubic interaction gives ${\cal{O}}(1)$ corrections to the background solution obtained for $g_3=0$ when  
\begin{equation}
\frac{\square\Phi}{\Lambda^3} \sim \mathcal{O}(1) \, .
\end{equation}
Solving for the scalar field far from the horizon we can neglect the galileon interaction: the field is sourced by the curvature squared term evaluated on the Schwarzschild solution
\begin{equation}
\square\Phi \sim \alpha \Mpl  \frac{r_s^2}{r^6} \, ;
\end{equation}
we can therefore estimate that at the horizon $r \sim r_s$
\begin{equation}
\frac{\square\Phi}{\Lambda^3} \sim  \frac{\alpha \Mpl}{r^4_s \Lambda^3} 
\end{equation}
that is $\mathcal{O}(1)$ for a black hole with $r_s \simeq 10\text{ km}$, using $\Lambda\simeq \Lambda_\alpha \simeq 10^{12} \text{km}^{-1}$ and $\Mpl\simeq 10^{38}\text{ km}^{-1}$.
In conclusion we obtained that, with a natural choice for scale of the cubic interaction, close to the horizon there are large corrections to the solution. In the next section we will solve perturbatively in $\tilde{\alpha} \ll 1$ the equation of motion of the action~\eqref{theory}. The perturbative expansion is schematically organized as follows,
\begin{align}
\Phi & \sim \Mpl \left(\tilde{\alpha}   +  \tilde{\alpha}^3 + \ldots   \right) \, ,
\\
h & \sim \tilde{\alpha}^2   +  \tilde{\alpha}^4 + \ldots   \, ,
\end{align}
where $h$ denotes the metric fluctuations with respect to the Schwarzschild background.

\subsubsection{Background solution}
\label{sec:bkgd}

In this section, we will  solve perturbatively the fields'   equations of motion up to  quadratic order in $\tilde{\alpha}$. The background metric can be parametrized as follows,
\begin{equation}
\D s^2 = - A(r) \D t^2  + \frac{\D r^2}{B(r)} + r^2  \left(  \D \theta^2 + \sin^2\theta \, \D\phi^2 \right) \, ,
\label{bkgd-metric-1}
\end{equation}
where, in full generality, we set the coefficient of $\D\Omega_{S^2}^2\equiv \D \theta^2 + \sin^2\theta \, \D\phi^2$ to $ r^2$, and where 
\begin{subequations}
\label{ABdef}
\begin{align}
A(r) & = 1- \frac{r_s}{r} + \tilde{\alpha}^2 \mathcal{A}(r)+ \mathcal{O}(\tilde{\alpha}^4) \, ,
\\
B(r) & = 1- \frac{r_s}{r} + \tilde{\alpha}^2 \mathcal{B}(r)+ \mathcal{O}(\tilde{\alpha}^4) \, .
\end{align}
\end{subequations}
The scalar field profile can be instead expanded as
\begin{equation}
\bar\Phi (r) = \Mpl \left( \tilde{\alpha} \,  \varphi(r)  + \mathcal{O}(\tilde{\alpha}^3) \right) \, .
\label{barPhi}
\end{equation}
Plugging into the Einstein equations, one can  find the equations for  $\mathcal{A}(r)$, $\mathcal{B}(r)$ and $\varphi(r)$. These do not usually admit a closed form solution if $g_3\neq 0$. However, integrating once the equation for $\varphi(r)$, one can find an expression for $\varphi'(r)$ in closed form:
\begin{equation}
\varphi'(r) = \frac{r^3-\sqrt{r^6 +16g_3 \left(4 r^3 r_s^3+ r^2 r_s^4+  r r_s^5-3  r_s^6 \right)}}{2g_3 r r_s^2 (4  r -3  r_s)} \, ,
\label{phip}
\end{equation}
where we fixed the integration constant in such a way that
$\varphi'(r)$ is finite at the horizon (which is sufficient to
guarantee $(\partial \varphi)^2$ is regular at the horizon) . Note that, in the limit $g_3\rightarrow0$, one recovers the result  of  \cite{Sotiriou:2014pfa},
\begin{equation}
\varphi'(r)\vert_{g_3\rightarrow0}  \rightarrow  -\frac{4 r_s \left(r^2+r r_s+r_s^2\right)}{r^4} \, .
\end{equation}
Regarding the equations for the metric components $\mathcal{A}(r)$ and $\mathcal{B}(r)$, it is convenient to look for solutions in the following form,
\begin{subequations}
\label{ABansatz}
\begin{align}
\mathcal{A}(r) & = \mathcal{B}(r) - \left(1-\frac{r_s}{r}\right) \int_r^\infty \D \rho \, \mathcal{F}_A(\rho) \, ,
\\
\mathcal{B}(r) & = \frac{1}{r} \int_{r_s}^r \D \rho \, \mathcal{F}_B(\rho) \, ,
\end{align}
\end{subequations}
where $\mathcal{F}_A$ and $\mathcal{F}_B$ are lengthy expressions, which  we do not write explicitly, that can be obtained straightforwardly by plugging \eqref{ABansatz} into the equations for $\mathcal{A}(r)$ and $\mathcal{B}(r)$.
The integrals in \eqref{ABansatz} can be shown to be convergent for all $r\in (r_s,\infty)$.\footnote{One can show that both $\mathcal{F}_A(\rho)$ and $\mathcal{F}_B(\rho)$ are continuous functions of $\rho$. In addition, they both approach a constant at $r=r_s$, while at large distances $\mathcal{F}_A(\rho\rightarrow\infty)\sim \frac{1}{\rho^3}$ and $\mathcal{F}_B(\rho\rightarrow\infty)\sim \frac{1}{\rho^2}$. }
As a sanity check, it can also be shown that $\mathcal{A}$ and $\mathcal{B}$ agree with eqs.~(3) and (4) of \cite{Pani:2011gy} in the limit $g_3\rightarrow0$.
In \eqref{ABansatz}, the boundary conditions have been chosen in such a way that the position of the horizon is still at $r=r_s$ (i.e., $\mathcal{A}$ and $\mathcal{B}$ both vanish at $r_s$).  Note that $\mathcal{A} \rightarrow \mathcal{B}$ as $r\rightarrow+\infty$, and they both scale as $\frac{1}{r}$. This implies a modification to the ADM mass with respect to general relativity and it is given by the formula
\begin{equation}
2GM  =  r_s - \tilde{\alpha}^2 \int_{r_s}^\infty \D \rho \, \mathcal{F}_B(\rho) \, ,
\end{equation}
where the correction can be easily computed numerically for fixed $g_3$ and $r_s$. 

Note that the scalar hair at large distances goes as (see \eqref{phip})
\begin{equation}
\varphi(r\rightarrow\infty) \rightarrow \frac{4r_s}{r} + \ldots
\end{equation}
Corrections in $g_3$ only enter starting from $\frac{1}{r^4}$. This means, for instance, that the scalar charge $P\equiv \frac{4\alpha}{r_s}$
 is unchanged with respect to the case with $g_3=0$ of \cite{Sotiriou:2014pfa}.

\subsubsection{Linearized equations for the perturbations and quasi-normal modes}
\label{sec:perts}

The additional operator in \eqref{theory} not only modifies the background geometry, but it also induces order-one corrections  to the linearized dynamics of the perturbations, as we  show in this section. 
To this end, we will use the following conventions for  the metric perturbations \cite{Franciolini:2018uyq,Franciolini:2018aad},
\begin{equation}
\delta g_{\mu\nu} = \delta g_{\mu\nu}^{\rm odd} + \delta g_{\mu\nu}^{\rm even} \, ,
\label{metricperts}
\end{equation}
with
\begin{equation} \label{odd metric perturbations version 0}
\delta g_{\mu\nu}^{\rm odd} = 
\begin{pmatrix}
0 & 0 & \varepsilon^k {}_j \nabla_k {\rm h_0}\\
0 & 0 & \varepsilon^k {}_j \nabla_k {\rm h_1}\\
\varepsilon^k {}_i \nabla_k {\rm h_0} \,\,\,\, &  \varepsilon^k {}_i \nabla_k {\rm
                                    h_1} & \,\,\,\, {1\over 2} (\varepsilon_i {}^k
                                           \nabla_k \nabla_j +
                                           \varepsilon_j {}^k \nabla_k
                                           \nabla_i ) {\rm h_2}
\end{pmatrix}
\end{equation}
which parametrizes parity-odd perturbations and where $\varepsilon_{ij}$ is defined in \eqref{gijLCt},\footnote{Indices are raised and lowered with $\gamma_{ij} \equiv {\rm diag}(1,\sin^{2}\theta)$.} and
\begin{equation}\label{even metric perturbations version 0}
\delta g_{\mu\nu}^{\rm even} =
\begin{pmatrix}
A H_0 & H_1 & \nabla_j {\mathcal H}_0 \\
H_1 & H_2 / B & \nabla_j {\mathcal H}_1 \\
\nabla_i{\mathcal H}_0 \,\,\,\, & \nabla_i {\mathcal H}_1 & \,\,\,\, C( {\cal K}
                                                   \gamma_{ij} + 
                                                   \nabla_i \nabla_j G)
\end{pmatrix} 
\end{equation}
for perturbations of the even type. $\nabla_{i}$ denotes a covariant derivative on the 2-sphere $S^{2}$, and in standard coordinates with metric $\gamma_{ij} \equiv {\rm diag}(1,\sin^{2}\theta)$  it follows that
\begin{eqnarray} \label{2 cov dev sphere}
\begin{split}
& \nabla_\theta \nabla_\theta = \partial_\theta^2 \,  , \qquad  \nabla_\phi \nabla_\phi
= \partial_\phi^2 + \sin \theta \cos \theta \partial_\theta 
\,  , 
\quad \nabla_\theta \nabla_\phi = \nabla_\phi \nabla_\theta
= \partial_\theta \partial_\phi - {\cos\theta \over \sin\theta} \partial_\phi \,
\\
& \nabla^2 = \partial_\theta^2 + {1\over \sin^2\theta} \partial_\phi^2 + {\cos\theta \over \sin\theta} \partial_\theta  \,  .
    \end{split}
\end{eqnarray}
The scalar perturbations are instead parametrized as $\Phi=\bar{\Phi}+\delta\Phi$, where $\bar \Phi$ is given in \eqref{barPhi}. Given the spherical symmetry of the background, it is guaranteed that even and odd perturbations do not couple at the level of the linearized equations of motion. For this reason, we can study the two sectors separately. In addition, the spherical symmetry and the time-translational invariance of the background allow to decompose a generic field perturbation ${\cal X}$ as ${\cal X}(t,r,\theta,\phi) = \sum_{\ell m}\int \frac{\D \omega}{2\pi} \, \e^{-i \omega t} {\cal X}_{\ell m}(\omega,r)Y_{\ell m}(\theta,\phi)$, where $Y_{\ell m}$ are the spherical harmonics and where ${\cal X}$ can be either  $\delta\Phi$ or any of the components in  $\delta g_{\mu\nu}$.

Let us start considering the odd sector, which is indeed the simplest
one, since it contains only one propagating  degree of
freedom. Following the standard procedure, we shall first choose the
Regge--Wheeler gauge, ${\rm h_2}=0$
\cite{Regge:1957td}.\footnote{Fixing the Regge--Wheeler gauge, ${\rm
    h_2}=0$,  at the level of the action \eqref{theory} is consistent
  as one does not lose any constraints (see, e.g.,
  \cite{Franciolini:2018uyq}).} Then,  using \eqref{odd metric
  perturbations version 0} and the spherical harmonic decomposition,
we can expand the action \eqref{theory} up to quadratic order in the
odd fields ${\rm h_0}(\omega,r)$ and ${\rm
  h_1}(\omega,r)$.\footnote{For simplicity, we will hereafter suppress
  $\ell$ and $m$ subscripts in  the spherical harmonic decomposition
  of the field perturbations, and also often altogether omit  the
  arguments in the fields, i.e.~${\cal X}_{\ell m}(\omega,r)$ will
  simply be often denoted as ${\cal X}(\omega,r)$, or just as ${\cal
    X}$.  That we are talking about fields decomposed in spherical
  harmonics and in frequency space should be clear from the context.} It is then straightforward to compute the equations of motion for ${\rm h_0}$ and ${\rm h_1}$, and find the combination that is algebraic in ${\rm h_0}$. Using this to  solve for ${\rm h_0}$ in terms of ${\rm h_1}$ and ${\rm h'_1}$, one arrives at a single equation for the propagating degree of freedom, which can be recast in the Schr\"odinger-like form
\begin{equation}
\frac{\D^2 Q}{\D r_\star^2} + \left( \omega^2 -V_Q\right) Q=0 \, ,
\label{cgeqQ}
\end{equation}
where the tortoise coordinate $r_\star$ is defined by
\begin{equation}
\frac{\D r_\star}{\D r} = \frac{1}{\sqrt{A B}} \, ,
\label{tortoiser}
\end{equation}
and where we introduced the field $Q$,  which is related to ${\rm h}_1$ via
\begin{equation}
{\rm h}_1(r) \equiv {\rm e}^{\mathcal{N}(r)}Q(r) \, ,
\label{cgeqQ2}
\end{equation}
The explicit expressions for $\mathcal{N}(r)$ and the potential $V_Q(r)$ are given in Appendix~\ref{app:cubicgalileonhairodd}. It is straightforward to check that, if $g_3\sim \mathcal{O}(1)$, the galileon coupling induces $\mathcal{O}(1)$ corrections to the potential $V_Q(r)$, with respect to \cite{Sotiriou:2014pfa}, at distances $r\sim r_s$, and therefore modifies the spectrum of quasi-normal modes.

The even sector is  more complicated, but can still be solved following, e.g., \cite{Franciolini:2018uyq}. The result is a system of coupled equations of the form
\begin{equation}
\frac{\D^2 \vec{\psi}}{\D r_\star^2} + \left( \omega^2 - \mathbb{V}_\psi\right) \vec{\psi}=0 \, ,
\label{cgeqpsieven}
\end{equation}
where $\vec{\psi}$ is a $2$-component vector, describing the two parity-even propagating degrees of freedom, and 
$\mathbb{V}_\psi$ is a $(2\times2)$-matrix. The relation between $\vec{\psi}$ and the perturbations $\delta\Phi$ and $\delta g_{\mu\nu}^{\rm even}$ (see eq.~\eqref{even metric perturbations version 0}),  the  explicit expression for $\mathbb{V}_\psi$, as well as the detailed procedure to obtain \eqref{cgeqpsieven}, are included in an ancillary notebook to the present paper. In the notebook, we also compute explicitly the eikonal limit for the potential  $\mathbb{V}_\psi$, which generalizes the result of \cite{Bryant:2021xdh}.

\subsection{Isospectrality and eikonal limit in the EFT for the perturbations}
\label{sec:recoveriso}

We  have shown above how  the presence of the cubic galileon operator modifies the properties of the hair generated by a sGB coupling.
This example illustrates the possibility of including many
other scalar operators in the theory. From this perspective, an
effective field theory is particularly useful, if one is interested in studying the dynamics of the perturbations in a  model-independent way.
This is the approach that was first introduced in \cite{Franciolini:2018uyq}  and that we will pursue  from now on.
Before generalizing the effective theory of \cite{Franciolini:2018uyq} to the case of slowly rotating black holes with scalar hair (see Section~\ref{sec:EFTsrbh} below), we  consider here another application of \cite{Franciolini:2018uyq} in the context of non-rotating black holes.

It is well known that the degeneracy between the even and odd QNM
spectra of general relativity is broken in the presence of a scalar
coupling to the Gauss--Bonnet operator. In the small-coupling limit,
for  the model of \cite{Sotiriou:2014pfa}, this has been shown
explicitly for low $\ell$'s in \cite{Blazquez-Salcedo:2016enn} and in
the eikonal limit in \cite{Bryant:2021xdh}. In this section, in the
spirit of studying how additional operators in the EFT can affect the
observables, we will show that the introduction of other operators can
have phenomenological implications, altering the spectrum of the
quasi-normal modes. In particular we will show explicitly that
suitable choices of the effective couplings allow to recover
isospectrality in the eikonal limit.\footnote{\label{footiso}By
  `isospectrality' we generically mean here that there is a degeneracy
  between even and odd sector at the level of the linearized equations
  of motion for the perturbations. Establishing whether or not there
  is an actual degeneracy at the level of the observed frequencies
  requires an additional piece of information, that we will not
  discuss in the present work, which is specifying how the scalar
  couples to the matter.} In this limit the equations simplify and the
conditions under which isospectrality is recovered can be stated
easily. The main point in this section is {\it not} to motivate the particular
choice of couplings giving rise to isospectrality, but rather to
illustrate, using the language of EFT, how a wide range of predictions is
possible (such as whether isospectrality is satisfied or not).

Let us take the EFT of \cite{Franciolini:2018uyq,Franciolini:2018aad} and let us focus on the following subset of operators,
\begin{equation}
S =  \int\D^4x \, \sqrt{-g} \left[
\frac{\Mpl^2}{2} R + \xi(r)\mathcal{G} -\Lambda(r) - f(r)g^{rr} 	+ M_2^4(r)(\delta g^{rr})^2
+ M_{13}^2(r) \delta g^{rr } \delta\hat{R} 
\right] \, ,
\label{EFT-ss0-GB}
\end{equation}
where $\xi(r)$ encodes the scalar coupling to the Gauss--Bonnet invariant $\mathcal{G}$, while $(\delta g^{rr})^2$ and $\delta g^{rr } \delta\hat{R} $ ($\hat{R} $ is the intrinsic curvature associated with constant-$r$ hypersurfaces  \cite{Franciolini:2018uyq}) are additional quadratic operators in the fields that can be generated in theories with derivative self-interactions for the scalar. Recall that, as opposed to  $M_2^4(r)$ and $M_{13}^2(r)$, which are in principle arbitrary, the coefficients $\Lambda(r)$ and $f(r)$ are fixed by the background equations of motion \cite{Franciolini:2018uyq}.  
Since the procedure to obtain the field equations has already been extensively discussed in \cite{Franciolini:2018uyq,Franciolini:2018aad} for non-rotating spacetimes, we will not repeat the derivation here, but we will simply highlight the main steps.

Instead of \eqref{bkgd-metric-1}, we find it more convenient to parametrize the background metric as follows,
\begin{equation}
\D s^2 = - A(r) \D t^2  + \frac{\D r^2}{A(r)} + C(r)  \left(  \D \theta^2 + \sin^2\theta \, \D\phi^2 \right) \, ,
\label{bkgd-metric-2}
\end{equation}
where we set $A=B$ and kept the coefficient $C$ of the line element on the $2$-sphere arbitrary. The relation between the tadpole coefficients  $f $ and $\Lambda$ and the background metric can be found by solving the background equations of motion \cite{Franciolini:2018uyq,Franciolini:2018aad}.
The equation for the odd mode can then be derived in complete analogy with Section~\ref{sec:perts}: we can choose the Regge--Wheeler gauge, ${\rm h}_2=0$, expand the action \eqref{EFT-ss0-GB} up to quadratic order in the metric perturbations \eqref{odd metric perturbations version 0} and integrate out ${\rm h}_0$ at the level of the linearized equations of motion. The final result is an equation of the form,
\begin{equation}
\frac{\D^2 Q}{\D r_\star^2} + \left( c_r^{-2}\omega^2 -V_Q\right) Q=0 \, ,
\label{oddeftstatic}
\end{equation}
 where now $\frac{\D r_\star}{\D r}=A^{-1}$ and $c_r$ plays the role of an effective speed of propagation (which can in general be non-unitary) in the radial direction. In the eikonal limit, $\ell\rightarrow\infty$, the function $c_r^{-2}$ and the potential $V_Q$ are
\begin{equation}
c_r^{-2} = \frac{\Mpl^2-4 A' \xi '-8 A \xi ''}{\Mpl^2-4 A' \xi '}\, ,
\qquad
V_Q  = \ell (\ell+1) \frac{ A \left(\Mpl^2 -4 A'\xi '-8 A\xi ''\right)}{\Mpl^2 C -4 A C' \xi '} + \mathcal{O}(\ell^0) \, ,
\label{crVQodd}
\end{equation}
where in $V_Q$ we neglected subleading terms in $1/\ell$. We can further assume that deviations from Schwarzschild are small, i.e., $A\equiv 1-\frac{r_s}{r}+ \epsilon  \mathcal{A}$ and $C\equiv r^2+ \epsilon \, \mathcal{C}$, where $\epsilon$ is some small coupling, and $\xi, \,  M_2^4, \,  M_{13}^2 \sim \mathcal{O}(\epsilon)$. Under this assumption, eqs.~\eqref{crVQodd} become
\begin{equation}
c_r^{-2} = 1-  \left(1-\frac{r_s}{r}\right) \frac{8    \xi ''}{\Mpl^2 } \, ,
\label{crlodd}
\end{equation}
\begin{equation}
V_Q  = \frac{\ell (\ell+1)}{r^2} \left(1-\frac{r_s}{r}\right) \left[ 1
+ \frac{\epsilon \mathcal{A}}{1-\frac{ r_s}{r}}
-\frac{\Mpl^2 \mathcal{C}-4 (2 r-3 r_s) \xi'-8 r (r_s-r) \xi''}{r^2\Mpl^2}
\right]\, .
\label{crVQodd-2}
\end{equation}
Following \cite{Franciolini:2018uyq,Franciolini:2018aad}, one can derive the equations governing the dynamics of the even modes. 
After the non-dynamical fields are integrated out on-shell and after a  suitable field redefinition that allows to remove first derivative terms in $r_\star$, the final system of equations  can be cast in the following form:
\begin{equation}
\frac{\D^2 \vec{\psi}}{\D r_\star^2} +  \mathbb{W}_\psi \cdot \vec{\psi}=0 \, ,
\label{cgeqpsieven-2}
\end{equation}
where $\mathbb{W}_\psi$ is a $(2\times2)$-matrix that depends on $r$, as well as on the frequency and the quantum number $\ell$. Note that the two equations will remain  in general coupled even in the eikonal limit, $\ell\rightarrow\infty$, as opposed to what happens in other situations, e.g., in the case of massive  and partially massless higher-spin fields in a Schwarzschild (de Sitter) spacetime \cite{Rosen:2020crj}. However, it is possible to choose the couplings $\xi$ $,  M_2^4$  and $  M_{13}^2$ in such a way that the two equations \eqref{cgeqpsieven-2} decouple. In the particular limit under consideration, this happens, for all $r$, if the following conditions are satisfied,\footnote{We stress that we are first taking the large-$\ell$ limit in \eqref{cgeqpsieven-2} and then expanding the result at linear order in $\epsilon$.}
\begin{subequations}
\label{MMtuning}
\begin{align}
M_2^4 & = -\frac{12 r_s \left(\left(-2 r^2+4 r r_s -3r_s^2\right) \xi'+r \left(2 r^2-5 r r_s +3 r_s^2\right) \xi''\right)}{r^3 (2 r-3 r_s)^2 (r-r_s)} \, ,
\\
M_{13}^2 & = -\frac{12 r_s \xi'}{2 r^2-3 r r_s} \, .
\end{align}
\end{subequations}
When eqs.~\eqref{MMtuning} hold, one can show not only that one of the off-diagonal entries of $\mathbb{W}_\psi$ becomes zero, but also that the corresponding equation becomes identical to  the odd equation \eqref{oddeftstatic}. In particular, the coefficients of the $\omega^2$ terms and the potentials equal \eqref{crlodd} and \eqref{crVQodd-2}, respectively.\footnote{The fact that the equations become identical is a result of the eikonal limit (the same happens for instance with the Regge--Wheeler and Zerilli equations in general relativity). Note that this is a sufficient condition for isospectrality, but it is not necessary. Different potentials can in fact lead to identical spectra, provided that they are related by a duality symmetry that preserves the boundary conditions \cite{Chandrasekhar:1985kt,Berti:2009kk,Glampedakis:2017rar,Rosen:2020crj}. } In this sense, one recovers `isospectrality' between the odd spectrum and one of the two even set of frequencies.\footnote{We stress, though, that the corresponding even mode is still in general a mixture of scalar and metric fluctuations (see also the comment in footnote \ref{footiso}). For analogous considerations in the case of scalar-tensor theories on exactly Schwarzschild background, see, e.g., \cite{Tattersall:2017erk}. }

Given the action \eqref{EFT-ss0-GB} with the coefficients
\eqref{MMtuning}, and given some educated guess on the form of the
scalar profile $\bar\Phi(r)$, one can in principle, by reverse
engineering, find explicit examples of covariant Lagrangians that
reduce to \eqref{EFT-ss0-GB} in unitary gauge.\footnote{See
  \cite{Pirtskhalava:2014esa} for a discussion, although in a
  different context, on how to obtain a reverse-engineered covariant
  model.} Although possible in general, this procedure does not
guarantee that the resulting effective scalar-tensor theory passes
extra sanity checks, such us having a well-defined expansion around a
Minkowski spacetime, or a consistent embedding in a unitary, causal
and local UV theory \cite{Adams:2006sv}.  Whether there exist explicit
examples that pass all these tests is left for future work. 

\section{Effective theory for perturbations of slowly rotating black holes}
\label{sec:EFTsrbh}

In this section, we generalize the results of \cite{Franciolini:2018uyq,Franciolini:2018aad} and construct an effective theory for perturbations around slowly rotating black holes with scalar hair. We assume here that parity is not broken and the scalar field is invariant under parity transformation --- we will discuss the case of parity-breaking theories in Section~\ref{sec:EFTCS}.
Even though what we will mostly have in mind (and refer to) below are asymptotically flat, slowly rotating black holes, we stress that our construction applies more generally to any stationary, axisymmetric, slowly rotating spacetime, whether or not it is asymptotically flat or has a horizon, including more exotic backgrounds (see \cite{Franciolini:2018aad} for an application of the EFT to wormholes in the non-rotating limit).

\subsection{General considerations and background metric}

In general relativity, a prominent example of stationary  and axisymmetric spacetime is given by  the Kerr metric, whose line element in  Boyer--Lindquist  coordinates is (see, e.g., \cite{Teukolsky:2014vca})
\begin{equation}
\begin{split}
\D s^{2} & = - \(1- \dfrac{r_s r}{\Sigma} \) \D t^{2} - \frac{2a  r_s r  }{\Sigma} \sin^{2}\theta \, \D t \, \D \phi
+ \frac{\Sigma}{\Delta}\D r^2 + \Sigma \, \D \theta^2
\\
&\qquad + \left( r^2 + a^2 + \frac{a^2 r_s r}{\Sigma} \sin^2\theta \right)\sin^2\theta \, \D\phi^2 \, ,
\end{split}
\end{equation}
where\footnote{The quantity $r_s$ is related to the  locations of the inner ($r_-$) and outer ($r_+$) horizons   by $r_\pm = \frac{r_s}{2}\pm \sqrt{\frac{r_s^2}{4}-a^2}$.}
\begin{equation}
\Sigma \equiv r^2 + a^2 \cos^2\theta \, ,
\qquad
\Delta \equiv r^2 - r_s r + a^2 \, ,
\end{equation}
and where $a$ is related to the angular momentum $J$ of the rotating black hole by
\begin{equation}
\label{spin}
J = M\, a.
\end{equation}
At linear order in the spin parameter $a$ the metric reduces to:
\begin{equation}
\label{kerrlina}
\D s^{2} = - \(1- \dfrac{r_s }{r} \) \D t^{2} +  \dfrac{\D r^{2}}{1-\frac{r_s}{r}} 
+ r^{2} \(\D \theta^{2} +\sin^{2}\theta \,\D \phi^{2}\)
- a \dfrac{2r_s}{r} \sin^{2}\theta\, \D t \, \D \phi   \, .
\end{equation}
In our discussion below, we wish to retain the interpretation of $a$  as the parameter related to the spin of the black hole by an equation similar to~\eqref{spin}. This leads us to consider metrics that, as \eqref{kerrlina},  are invariant under the transformations $\{t \rightarrow -t , a \rightarrow -a\}$ and $\{t \rightarrow -t , \phi \rightarrow -\phi\}$. These symmetries, together with the assumption of a smooth $a\rightarrow 0$ limit, imply that the only term linear in $a$ is in the $(t,\phi)$ component of the metric.
Motivated by this observation and the form of the components in  \eqref{kerrlina},  we will  focus below on  the following class of  stationary, axisymmetric background metrics:
\begin{equation} \label{eq:background}
\D s^{2} = - A(r) \D t^{2} + \dfrac{1}{B(r)} \D r^{2} + C(r) \(\D \theta^{2} +\sin^{2}\theta \, \D \phi^{2}\)  - 2 \,a \, D(r) \sin^{2}\theta\, \D t \,  \D \phi \, ,
\end{equation}
where $A(r)$, $B(r)$, $C(r)$ and $D(r)$ are arbitrary functions of $r$.\footnote{In full generality, we are allowed, by the possibility of redefining the radial coordinate $r$, to fix one of the four functions $A(r)$, $B(r)$, $C(r)$, $D(r)$.} 

One might wonder how general the ansatz \eqref{eq:background} is.
As shown in Refs.~\cite{Johannsen:2013szh,Carson:2020dez},
eq.~\eqref{eq:background} represents, at linear order in $a$, the most
general ansatz for a stationary, axisymmetric and asymptotically flat
black hole solution  with separable geodesic equations, that preserves
the Kerr symmetries and reduces to a spherically symmetric background
in the limit $a\rightarrow 0$. Note that an overall $\theta$-dependent
conformal factor could  in principle be included in
\eqref{eq:background} while still preserving the Kerr
symmetries~\cite{Carson:2020dez}. In that case, in the limit
$a\rightarrow 0$, one would recover a metric that is  conformally
related to the Schwarzschild solution.  In what follows, we will for
simplicity disregard  this possibility, although such a
$\theta$-dependent conformal factor  could be straightforwardly
accounted for in the construction of the EFT.\footnote{Such a
  $\theta$-dependent conformal factor is strongly constrained by
  weak-field solar system observations, so that we can safely neglect
  it~\cite{Johannsen:2013szh,Carson:2020dez}.} Similarly, we will not
consider here the possibility of more general rotating metrics with
arbitrary $\theta$-dependence in the components. As an additional
motivation for our ansatz \eqref{eq:background}, we stress that
\eqref{eq:background} is a good solution in many explicit examples of
slowly rotating hairy black holes (see for instance Refs.~\cite{Pani:2011gy,Bakopoulos:2021dry} and Appendix~\ref{app:example} below).

Given the ansatz \eqref{eq:background}, one still needs to specify the form of the scalar  profile that sources it. This will be relevant when we construct the EFT for the perturbations in the unitary gauge. 
Note that, although the scalar's energy-momentum tensor is expected to have the same symmetries of the background metric, this need not be the case, in principle, for the scalar $\bar\Phi$ itself. A simple example of this is given by shift-symmetric theories where the solution for $\bar \Phi$ contains a linear term in time, $\bar{\Phi}\equiv t +\varphi(r)$ (see, e.g., \cite{Babichev:2013cya,Babichev:2017guv,Minamitsuji:2018vuw,BenAchour:2020wiw}). Despite the explicit time dependence in the field, the time-translational invariance of the system follows from the shift invariance of the scalar action.\footnote{There is in fact a diagonal combination of time translations and constant shifts  that is still linearly realized on $\Phi$.} In this section, we work under the  simplifying assumption that  the scalar background for $\Phi$ inherits the same symmetries of the metric \eqref{eq:background}.
As a consequence, $\bar \Phi$ cannot depend explicitly  on $t$, nor on $\phi$. In addition, invariance under $\{ a \rightarrow -a, \phi \rightarrow -\phi\}$ forbids  linear terms in $a$ in $\bar \Phi$. 
Moreover, we will assume that $\Phi$ is a scalar under parity,
i.e.~that it is invariant under
$\{\theta\rightarrow\pi-\theta,\phi\rightarrow\phi+\pi\}$, so that
parity is  broken neither explicitly in the action nor spontaneously
by the background. This tells us  that $\bar{\Phi}$ cannot depend on
$\theta$  (in a way that solves non-trivially the background Einstein
equations). As a result, $\bar{\Phi}$ must be a  function of $r$ only
($\bar{\Phi}\equiv\bar \Phi(r)$) and with no dependence on $a$ at
linear order \cite{Sotiriou:2014pfa}. We will partially relax these
assumptions in Section~\ref{sec:EFTCS}, where we will allow the
breaking of parity (in such a way that $\bar{\Phi}$ can depend on
$\theta$) and the scalar $\bar{\Phi}$ to not be invariant under $\{ a
\rightarrow -a, \phi \rightarrow -\phi\}$.\footnote{It would also be interesting to study more general cases and generalize the EFT to  perturbations around solutions such that $\bar{\Phi}\equiv t +\varphi(r)$. This is left for future work.}

\subsection{Effective theory for the perturbations}
\label{subsec:EFTperts}

To construct the EFT for the perturbations, we will work in the unitary gauge, defined by $\delta \Phi \equiv 0$. Since the scalar background $\bar{\Phi}$ depends on $r$, the unitary gauge breaks radial diffeomorphisms. This suggests the introduction of a radial foliation with hypersurfaces  satisfying $\bar{\Phi} = \rm constant$, provided that the scalar profile is non-trivial, i.e., $\nabla_{\mu}\bar{\Phi} \neq 0$, and that the original map $\bar{\Phi}: r \to \bar{\Phi}(r)$ is an injective map on the whole domain in which the radial coordinate $r$ is defined. The continuity of $\bar{\Phi}(r)$, together with the previous conditions, implies that $\bar{\Phi}(r)$ has to be a  strictly monotonic function  in order to have a well-defined foliation.

After choosing the unitary gauge and fixing the radial foliation, the effective action will be the most general action invariant, up to boundary terms, under the residual temporal and angular diffeomorphisms.
In analogy with \cite{Franciolini:2018uyq,Franciolini:2018aad}, the geometric objects that constitute the basic building blocks are
$g_{\mu\nu}$, $R_{\mu\nu\rho\sigma}$, $\nabla_{\mu}$, 
$g^{rr}$ and the extrinsic curvature $K_{\mu\nu}$, with  coefficients that can be arbitrary functions of $r$.
For more details about the notation and the geometric decomposition, see \cite{Franciolini:2018uyq} and Appendices \ref{app:foliation} and \ref{app:infvar} below.

\subsubsection{Linear order}
\label{sec:tadpoles}

In this section, we focus on the terms of the EFT that are linear in the field perturbations and find a set of  independent operators that form a basis.
 
At linear order in perturbations, the most general action is
\begin{equation}
\label{eq:tadpoles}
S_{\rm tadpoles} = \int \D^{4}x \sqrt{- g} \[ - \Lambda(r) - f(r) \delta g^{rr} + k^{\mu\nu}(r)\delta K_{\mu\nu}+ \xi^{\mu\nu\alpha\beta}(r) \delta R_{\mu\nu\alpha\beta}
\] \, ,
\end{equation}
with $\Lambda(r)$, $f(r)$, $k^{\mu\nu}(r)$, $\xi^{\mu\nu\alpha\beta}(r)$ arbitrary functions of the background metric and its derivatives. 

Let us start considering the tadpole $k^{\mu\nu}(r) K_{\mu\nu}$. The orthogonality condition~\eqref{eq:ort_k} implies that the only non-zero part is $k^{ab}(r) K_{ab}$. The function $k^{ab}$ is secretly a function of the background metric and thus inherits its structure. Therefore we can write:
\begin{equation}
k^{ab}(r) \delta K_{ab} =k^{tt}(r) \delta K_{tt} + k^{\theta\theta}(r) \( \delta K_{\theta\theta} + \dfrac{\delta K_{\phi\phi}}{\sin^{2}\theta} \) + a \, k^{t\phi}(r) \frac{\delta K_{t\phi}}{\sin^2\theta}.
\label{kappaab}
\end{equation}
Using the fact that the trace of $K=\nabla_{\mu}n^{\mu}$ is a total derivative and can be recast, up to a total derivative, as the $\Lambda(r)$ and $f(r)g^{rr}$ terms, we can get rid of one of the components in \eqref{kappaab}. For instance we can add a term $- k^{\theta \theta}(r) C(r) K$ and get rid of the second term  in \eqref{kappaab}.
We are left with two terms, which can be conveniently expressed as
\begin{equation}
k^{ab}(r) K_{ab} \equiv \alpha (r) \bar{K}_{\mu\nu} \delta K^{\mu\nu} + a \, \sin^2\theta \,  \beta (r) \delta K^{t\phi}.
\label{kappaab-2}
\end{equation}
The first term in \eqref{kappaab-2} was already present in the EFT for non-rotating black holes \cite{Franciolini:2018uyq}. The second term, proportional to the spin parameter $a$, is instead new and accounts for the fact that the  spacetime is `less symmetric', compared to the non-rotating case, which translates into a larger set of independent operators in the EFT. 

Let us now consider the tadpole $\xi^{\mu\nu\alpha\beta}(r) R_{\mu\nu\alpha\beta}$.
As before, $\xi^{\mu\nu\alpha\beta}(r)$ is secretly a function of the background metric and its derivatives, which dictate therefore its tensor structure; moreover it has the same symmetry properties as $R_{\mu\nu\alpha\beta}$. This fact, together with the structure of the background metric, implies that out of the 20 components of the Riemann tensor, only 6 could give rise to new tadpoles:
\begin{equation}
R_{rtrt}, \quad (R_{\theta r \theta r} + R_{\phi r \phi r}), \quad (R_{\theta t \theta t} + R_{\phi t \phi t}), \quad R_{\theta\phi\theta\phi}, \quad R_{rtr\phi}, \quad R_{\theta t \theta\phi} \, .
\label{tadsR}
\end{equation}
We will show  in fact that there is only one independent operator, corresponding to the Ricci scalar $R$.
The first four terms in \eqref{tadsR} are the same as those appearing in \cite{Franciolini:2018uyq}  for non-rotating spacetimes. Three of them can be re-expressed in terms of  $R$, the other tadpoles or boundary terms---see Appendix A of~\cite{Franciolini:2018uyq}.
The last two tadpoles $R_{rtr\phi}$, $R_{\theta t \theta\phi}$ can instead be dealt with as follows. Since they both have $t\phi$ indices and are symmetric in their exchange, they will necessarily be contracted with $\bar{g}_{t\phi}$ or its derivatives, which is already order-$a$. Since we are interested in the EFT  at linear order in $a$, we are thus allowed to think of $R_{rtr\phi}$ and $R_{\theta t \theta\phi}$ as being computed at the leading order $a=0$. Then, from eq.~\eqref{eq:riemann_variation} we can read off  the variation of the Riemann tensor at the order  $a=0$, that is
\begin{equation}
\begin{split}
\delta R_{\theta t \theta \phi} &= \bar{g}_{\theta\theta } \(\nabla_{\theta} \delta \Gamma_{t\phi}^{\theta}- \nabla_{\phi} \delta \Gamma_{t\theta}^{\theta} \), \\
\delta R_{r t r\phi} &= \bar{g}_{rr} \(\nabla_{r} \delta \Gamma_{t\phi}^{r}- \nabla_{\phi} \delta \Gamma_{tr}^{r} \).
\end{split}
\end{equation}
The terms with covariant derivatives in $\theta$ and $\phi$ can be eliminated, up to boundary terms, by integrations by parts, while the term with an $r$ derivative can be also expressed as a $t\phi$ tadpole by using equation~\eqref{eq:KGamma_tphi}.

As a result, the most general tadpole action boils down to
\begin{equation}
\label{eq:tadpoles-2}
S_{\rm tadpoles} = \int \D^{4}x \sqrt{- g} \[\frac{1}{2}M_1^2(r)R  - \Lambda(r) - f(r)  g^{rr} - \alpha(r) \bar K^{\mu\nu} K_{\mu\nu}- a \, \sin^2\theta \, \beta(r)   g^{t\phi} 
\] \, ,
\end{equation}
where, for convenience, we used $g^{t\phi}$ instead of $K^{t\phi}$.\footnote{The two operators are equivalent up to integrations by parts---see Appendix~\ref{appendix:tphi_tadpole}.}

The coefficients in \eqref{eq:tadpoles-2} are not all arbitrary. We report the form of the background equations in the case where $M_1\equiv \Mpl$ is taken to be constant. The $tt$ and $rr$-Einstein equations fix $f(r)$ and $\Lambda(r)$:\footnote{Note that eqs.~\eqref{tadpolefL} and \eqref{tadpoleeqalpha} here  match eqs.~(2.11)-(2.13) of \cite{Franciolini:2018uyq}, which are valid in the non-rotating limit $a=0$.}
\begin{subequations}
 \label{tadpolefL}
\begin{align}
f(r) & = \left( \frac{A'C'}{4 AC}-\frac{B'C'}{4 BC}+\frac{C'^2}{4C^2}-\frac{C''}{2C} \right) \Mpl^2
\nonumber
\\
&\qquad
+ \left( \frac{A' B'}{AB}+\frac{2 A'C'}{AC}-\frac{4 A'^2}{A^2}+\frac{2 A''}{A}-\frac{2 C'^2}{C^2}  \right) \frac{\alpha}{8}
 + \frac{A' }{4 A}\alpha' \, ,
 \label{tadpolef}
\\
\Lambda(r)& = \left( -\frac{B A'C'}{4 AC}-\frac{B'C'}{4 C}-\frac{BC''}{2 C}+\frac{1}{C} \right) \Mpl^2
\nonumber
\\
&\qquad
+ B \left( \frac{A' B'}{A B}+\frac{2 A'C'}{AC}-\frac{4 A'^2}{A^2}+\frac{2 A''}{A}-\frac{2 C'^2}{C^2}  \right) \frac{\alpha}{8}
 + \frac{B A'}{4 A} \alpha' \, .
 \label{tadpoleL}
\end{align}
\end{subequations}
In addition, the $t\phi$-component of the background Einstein equations can be used to derive an expression for $\beta (r)$:
\begin{align}
\beta(r) & = BD \left(-\frac{A' B'}{4 AB}-\frac{A'D'}{4 AD}+\frac{A'^2}{4 A^2}-\frac{A''}{2 A}+\frac{B'D'}{4 BD}-\frac{1}{BC}+\frac{D''}{2 D} \right) \Mpl^2
\nonumber
\\
&\qquad
+  BD \left( -\frac{A' B'}{4 A B}+\frac{A'C'}{2 AC}-\frac{3 A'D'}{4 AD}+\frac{3 A'^2}{4 A^2}-\frac{A''}{2 A}+\frac{B'D'}{4 BD}-\frac{C'D'}{2 CD}+\frac{D''}{2D}  \right) \alpha
\nonumber
\\
&\qquad
 +  BD \left( \frac{D'}{D}-\frac{A'}{A} \right) \frac{\alpha'}{2} \, ,
\end{align}
while the $\theta\theta$ (or $\phi\phi$) equation yields the following differential equation for $\alpha(r)$:
\begin{multline}
\left( \frac{A'}{A}-\frac{C'}{C} \right) \alpha'
+ \left( \frac{A' B'}{2 A B}+\frac{A'C'}{2 AC}-\frac{3 A'^2}{2 A^2}+\frac{A''}{A}-\frac{B'C'}{2 BC}+\frac{C'^2}{C^2}-\frac{C''}{C} \right) \alpha
\\
+ \left( \frac{A' B'}{2 A B}+\frac{A'C'}{2 AC}-\frac{A'^2}{2 A^2}+\frac{A''}{A}-\frac{B'C'}{2 BC}+\frac{2}{BC}-\frac{C''}{C} \right) \Mpl^2 =0 \, .
\label{tadpoleeqalpha}
\end{multline}

In the subset of theories in which $\alpha(r)=\beta(r)\equiv 0$---corresponding to theories with no higher derivative operators---the background equations simplify and reduce to those of Ref.~\cite{Franciolini:2018uyq}, plus a self-consistency differential equation that relates the function $D(r)$ to the other background functions. Choosing coordinates in which $A(r)=B(r)$, this equation takes a particularly simple form:
\begin{equation}
\label{eq:tadpoleD}
D''(r) C(r) - D(r) C''(r)=0.
\end{equation}
Using this equation it is easy to generalise known examples of spherically symmetric hairy black holes to slowly rotating ones.
An explicit realization within this type of theories is given in Appendix~\ref{app:example}.

Since the most studied example of scalar hair in the literature is the one sourced by a coupling to the Gauss--Bonnet invariant, it might be sometimes  convenient to isolate the Gauss--Bonnet operator, as we did already in eqs.~\eqref{theory} and~\eqref{EFT-ss0-GB}:
\begin{equation}
\begin{split}
S_{\rm tad.} = \int \D^{4}x \sqrt{- g} \[\frac{1}{2}M_1^2(r)R + \xi(r) \mathcal{G} - \Lambda(r) - f(r)  g^{rr} - \alpha(r) \bar K^{\mu\nu} K_{\mu\nu}- a \, \sin^2\theta \, \beta(r)   g^{t\phi} 
\].
\end{split}
\label{eq:tadpoles-3}
\end{equation}
With this choice there is clearly a redundancy in the effective Lagrangian at linear order in perturbations, because not all the operators in \eqref{eq:tadpoles-3} are independent, as we showed above. However, if the goal is to apply the EFT to study perturbations around  black hole solutions with scalar hair, the choice \eqref{eq:tadpoles-3} is particularly convenient.
It allows, indeed, to more easily perform the matching with explicit models that involve a coupling to the Gauss--Bonnet invariant, which is one of the most studied examples in the literature  evading the no-hair theorem. In fact, Ref.~\cite{Creminelli:2020lxn} showed that in the context of scalar-tensor theories with no ghost degrees of freedom the no-hair theorem of Hui and Nicolis~\cite{Hui:2012qt} can be evaded only in the presence of the operator $\Phi\mathcal{G}$, that is a linear coupling of the scalar to the Gauss--Bonnet invariant. In this sense, the EFT is capturing all possible corrections to the Einstein--Hilbert action \textit{and} the sGB coupling.\footnote{The fact of writing explicitly  the Gauss--Bonnet operator in the effective action \eqref{eq:tadpoles-3} should not be surprising. Note that we always do the same with the Einstein--Hilbert term. Given the special role played in the theory, it is just a matter of convenience to write it fully non-linearly, instead of expanding it in perturbations.}

The general set of background equations in this parametrization and
with all the tadpoles included is quite cumbersome. We therefore do not write them here, but make the derivation and result available in an ancillary notebook. 

\subsubsection{Quadratic action} \label{sec:quadratic_action}

In this section, we extend the action \eqref{eq:tadpoles-3} to higher orders in the fields. In particular, we introduce quadratic operators, which we will use to study the linearized dynamics of the perturbations around black hole solutions with hair. We will omit the derivation, which follows the one in \cite{Franciolini:2018uyq},  of the complete set of operators that enter at quadratic order in the EFT. Instead, we will focus on a particular example and show how the linearized dynamics gets modified in the presence of the considered operator.  

The linearized field equations on static spherically symmetric spacetimes are in general amenable to a separation of variables and a decomposition in spherical harmonics. For axisymmetric rotating backgrounds,  it is in many cases also  possible  to decompose the fields in radial and angular components, provided that the spherical harmonics are suitably replaced by (spin-weighted) spheroidal harmonics. This is for instance the case of  massless  perturbations of generic spin $s$ on a Kerr background \cite{Teukolsky:1973ha,Press:1973zz}. However, at the linear order in the black hole spin parameter $a$, the situation is much simpler: the ansatz \eqref{metricperts}, with $\delta g_{\mu\nu}^{\rm odd}$ and  $\delta g_{\mu\nu}^{\rm even}$ given in \eqref{odd metric perturbations version 0} and \eqref{even metric perturbations version 0}, remains a good parametrization of the metric perturbations; in addition, since the fields defining  the components of  $\delta g_{\mu\nu}$ in \eqref{odd metric perturbations version 0} and \eqref{even metric perturbations version 0} transform as scalars, they can still be decomposed in spherical harmonics, in complete analogy with the non-rotating case ($a=0$).
The only main difference is that the linearized equations for $\delta g_{\mu\nu}^{\rm odd}$ and  $\delta g_{\mu\nu}^{\rm even}$ will no longer be decoupled, with mixing terms between even and odd components appearing at linear order in $a$ (see, e.g., \cite{Pani:2013pma}). We refer the reader to appendices~\ref{app:spherical_harmonics} and~\ref{app:gauge} for a detailed discussion of this decomposition and our choice of gauge.
In the following, we derive the coupled equations for our EFT at the leading order in derivatives, showing that the result correctly reproduces the non-rotating case studied in \cite{Franciolini:2018uyq,Franciolini:2018aad}. 

The simplest non-trivial example of quadratic operator is $(\delta g^{rr})^2$. At the leading order in the derivative expansion, the EFT takes on the following simple form:\footnote{An explicit model belonging to the class of theories described by \eqref{EFT0der} with $M_2^4=0$ is discussed in Appendix~\ref{app:example}. Other examples captured by  \eqref{EFT0der}  with $M_2^4=0$ are also discussed in \cite{Herdeiro:2015waa,Bakopoulos:2021dry}.}
\begin{equation}
 S =  \int\D^4x \, \sqrt{-g} \left[
\frac{1}{2}\Mpl^2 R -\Lambda(r) - f(r)g^{rr} 
	+ M_2^4(r)(\delta g^{rr})^2
\right] \, .
\label{EFT0der}
\end{equation}
To obtain the quadratic action  for the  propagating degrees of freedom (for generic $\ell$, one scalar and the two graviton polarizations),  we shall proceed as follows. We first  parametrize the fluctuations of the metric as in \eqref{metricperts}, \eqref{odd metric perturbations version 0} and  \eqref{even metric perturbations version 0}; the residual gauge freedom allows us to get rid of $3$ out of the $10$ independent field variables; we plug \eqref{metricperts} into \eqref{EFT0der} and expand up to quadratic order in the fields; we `integrate out' the $4$ non-dynamical fields and compute the $3$ remaining equations for the propagating degrees of freedom. The derivation of the general set of coupled equations in the case of \eqref{EFT0der} is outlined in Appendix~\ref{app:eoms}.
Here, we emphasize that, as we discussed above and it is clear from Appendix~\ref{app:eoms}, the linearized equations now mix even and odd modes. In particular, the equations of the even modes with fixed $\ell$ contain a  coupling to the odd field with $\ell\pm1$, and vice versa \cite{1991RSPSA.433..423C}.

Since the mixing terms between odd and even modes are of order $\mathcal{O}(a)$, they generate corrections to the quasi-normal spectrum at quadratic order in $a$. Therefore, if one is interested in computing the QNMs at linear order in the spin parameter, these mixing terms can be neglected and the equations simplify considerably~\cite{10.1143/ptp/90.5.977,Pani:2012bp}. In this spirit, we show now the form of the odd equation if we neglect the coupling to the even modes.  For simplicity let us set $B=A$ in \eqref{eq:background}  and let us fix the gauge where ${\rm h}_2=0 $ in \eqref{odd metric perturbations version 0} (see appendix~\ref{app:gauge} for more details). As it is clear from \eqref{odd metric perturbations version 0}, the coefficient $M_2^4$ will not enter the dynamics of the odd sector.  Following the procedure outlined in Appendix~\ref{app:eoms}, after straightforward  manipulations, we arrive at the following equation for the axial mode:
\begin{equation}
\frac{\D^2 Q}{\D r_\star^2} + \left( \omega^2 -\frac{2 a m  \omega  D}{C} -V_Q\right) Q=0 \, ,
\label{oddeftgkerr}
\end{equation}
where  $\frac{\D r_\star}{\D r}=A^{-1}=B^{-1}$ and where we defined
\begin{equation}
{\rm h}_1 \equiv \frac{\sqrt{C}}{A} \left(1 -\frac{a m D}{\omega C} \right) Q \, .
\end{equation}
The potential $V_Q$ is given by
\begin{multline}
V_Q =   \frac{4 \left(\ell^2+\ell-3\right)AC-2A C A'C'-2A C^2 A''+3 A^2 C'^2}{4C^2} 
\\
+ \frac{2 a m A \left(C A'-3 AC'\right) \left(CD'-DC'\right)}{\ell (\ell+1) \omega  C^3}\, .
\end{multline}
Note that, in the limit of a slowly rotating Kerr spacetime, i.e.~$A=B=1-\frac{r_s}{r}$, $C= r^2$ and $D = \frac{r_s}{r}$, we recover precisely the equation of motion of the axial mode in general relativity (see, e.g., eq.~(115) of \cite{Pani:2013pma}).

\section{Effective theory with parity-breaking operators}
\label{sec:EFTCS}

In Section~\ref{sec:EFTsrbh} we showed how to construct the effective theory for the perturbations of slowly rotating hairy black holes under the assumption that parity is not broken. It is instructive to see how the previous conclusions change and what the EFT looks like for black hole solutions whose hair is sourced by a field profile that is not invariant under parity transformations. One notable example in this class of theories is given by dynamical Chern--Simons gravity, where the (psuedo) scalar $\Phi$ couples linearly to the Pontryagin density $\prescript{*}{}{\! R}^{\mu\nu\rho\sigma}R_{\mu\nu\rho\sigma}\equiv \frac{1}{2} {\varepsilon^{\mu\nu}}_{\lambda\tau}R^{\lambda\tau\rho\sigma}R_{\mu\nu\rho\sigma}$ \cite{Jackiw:2003pm,Alexander:2009tp}. Both static and rotating black hole solutions in dynamical Chern--Simons gravity, as well as the analysis of the perturbations, have been widely discussed in the literature \cite{Smith:2007jm, Grumiller:2007rv, Yunes:2007ss,  Yunes:2009hc,Konno:2009kg, Cardoso:2009pk, Molina:2010fb, Pani:2011gy, Yagi:2012ya, Kimura:2018nxk,Wagle:2021tam,Srivastava:2021imr}. In this section, we extend the effective framework introduced in Section~\ref{sec:EFTsrbh} in such a way to capture, in a model-independent way, also modifications of gravity induced by couplings to  parity-odd operators, like the Chern--Simons one. The only assumption we will need is that of a well-defined radial foliation, that is: for every fixed value of the angular coordinate $\theta_{\star}$, the scalar background $\bar{\Phi}(r,\theta_{\star})$ needs to be a strongly monotonic function of $r$ defined on the whole domain of definition of the radial coordinate $r$.
The EFT will allow us to describe, in a single framework, many cases previously studied in the literature, as well as more general situations that have not been discussed before. As we will see in a moment, the purely dynamical Chern--Simons gravity with no additional operators is, however, an exception.

To construct the EFT for the perturbations, we assume again the parametrization \eqref{eq:background} for the background metric. Even though its form  is unchanged with respect to the case considered in Section~\ref{sec:EFTsrbh}, we now allow the background field $\bar \Phi $ to acquire some dependence on the angle $\theta$. Assuming that in the zero spin limit, $a=0$, one recovers the results of \cite{Franciolini:2018uyq}, the $\theta$-dependence will always be proportional to $a$. Thus, we can postulate the following generic profile for the background of $\Phi$,
\begin{equation}
\bar \Phi (r,\theta) = \bar \Phi_0(r) + a \,  \bar \Phi_1(r,\theta) \, .
\label{eq:phibarCS}
\end{equation}
The form of $\Phi_1$ as function of $\theta$ depends on the model and is in general determined by the background equations of motion. In the case of  dynamical Chern--Simons gravity (with no additional operators) the scalar background takes the simple form $\bar \Phi (r,\theta)\vert_{\rm dCS}=a \, f_{\rm dCS}(r) \cos \theta$. This implies $\bar \Phi (r,\pi/2)\vert_{\rm dCS}\equiv~0$, so that dynamical Chern--Simons gravity turns out to be a pathological example for our construction, since the scalar field cannot be used to define a radial foliation.
In more general theories, however, one expects additional operators beyond the Chern--Simons one, and in general $\Phi_{0}(r)\neq 0$.\footnote{When considering the EFT in the narrower context of hairy black holes in shift-symmetric scalar-tensor theories, a scalar profile with $\Phi_{0}\neq 0$ and $\Phi_{1} \neq 0$ requires the presence of a \textit{linear} coupling of the scalar field to both the Gauss--Bonnet invariant and the Pontryagin density~\cite{Creminelli:2020lxn}.} Explicit examples of slowly rotating black hole solutions in theories that include a coupling to the Pontryagin density plus additional operators are discussed e.g.~in \cite{Pani:2011gy}. In these examples both the metric and the scalar background profiles are of the form of our ansatz \eqref{eq:background} and \eqref{eq:phibarCS}.

Given the background profiles \eqref{eq:background} and \eqref{eq:phibarCS}, we would like now to generalize the construction of the EFT presented in Section~\ref{sec:EFTsrbh}. 
In addition to  $g_{\mu\nu}$, $R_{\mu\nu\rho\sigma}$, $K_{\mu\nu}$, $\nabla_{\mu}$  and $g^{rr}$ that constitute the building blocks of the EFT  in Section~\ref{sec:EFTsrbh}, we are now allowed to contract spacetime indices also with the totally antisymmetric tensor $\varepsilon^{\mu\nu\rho\sigma}$. This gives more freedom and extend the number of independent operators in the EFT at each order in perturbation theory. The construction of the Lagrangian for the tadpoles closely follows the logic in Section~\ref{sec:tadpoles}. One can similarly fix the unitary gauge by fixing the $r$-diffeomorphisms in such a way to remove the $\delta \Phi$ fluctuations altogether. The main difference in the construction of the foliation is that now the unit vector $n_\mu$, defined in \eqref{nmu}, will have also a nonzero $\theta$-component at linear order in $a$, due to the non-trivial $\theta$-dependence in $\bar \Phi$.
This implies, for instance, that the extrinsic curvature $K_{\mu\nu}$ can have now nonzero contravariant $r$-components at linear order in $a$ (see eqs.~\eqref{eq:ext_curv} and \eqref{eq:ort_k}). This modifies some of the steps  in  Section~\ref{sec:tadpoles}, making the derivation of the tadpole Lagrangian slightly more complicated.
Thus, instead of following the same logic, we provide here a simpler way to count the number of independent operators in the EFT at the linear order in perturbations. In the unitary gauge, the most general theory of a black hole with (pseudo-)scalar hair \eqref{eq:phibarCS} can be now written in the form 
\begin{equation}
S = \int\D^4 x\sqrt{-g} \, \mathcal{L}\left(g_{\mu\nu}, \varepsilon^{\mu\nu\lambda\rho}, 
R_{\mu\nu\alpha\beta}, g^{rr}, g^{r\theta} , K_{\mu\nu} ,\nabla_\mu ; r , \theta 
\right) \, ,
\label{Tintrob}
\end{equation}
where the dependence on $g^{r\theta}$ and $\theta $ follows from the new $\theta$-dependent profile \eqref{eq:phibarCS}. When expanded in perturbations, up to integrations by parts, the Lagrangian at linear order can be in general recast in the form
\begin{equation}
\mathcal{L}_\text{tadpoles} \sim F_{\mu\nu}[\partial_r,\partial_\theta , \bar{g}, \varepsilon;r,\theta] \delta g^{\mu\nu} \, ,
\label{tadpolesFdeltag}
\end{equation}
where the matrix $F_{\mu\nu}$ inherits its structure from  the ingredients in \eqref{Tintrob} suitably contracted and computed on the background, and from the integrations by parts. In other words, it can be thought of as being some generic (symmetric) matrix resulting from certain contractions among the tensor $\varepsilon^{\mu\nu\lambda\rho}$, the spacetime derivatives $\partial_\mu$ projected on the $r$ and $\theta$-directions, and a certain number of  background metric tensors $\bar{g}$.
Thus, counting the number of independent tadpoles in \eqref{tadpolesFdeltag}  amounts in the end to finding the number of independent components in $F_{\mu\nu}$.\footnote{Of course, this logic reproduces the counting of operators in simpler EFTs based on a similar construction, such as~\cite{Franciolini:2018uyq}.}
By definition,  $\varepsilon^{\mu\nu\lambda\rho}$ is nonzero only when
its indices are all different. In addition, since the background is
independent of $t$ and $\phi$ by construction, no $t$ or
$\phi$-derivatives can appear in $F$.  Thus, given the form of
$\bar{g}_{\mu\nu}$ in \eqref{eq:background}, it follows that the only metric fluctuations in \eqref{tadpolesFdeltag} that allow to contract the Lorentz indices in a consistent way are
\begin{equation}
\mathcal{L}_\text{tadpoles} \supseteq \delta g^{tt}, \,  \delta g^{rr}, \, \left( \delta g^{\theta\theta}+ \sin^2\theta \,  \delta g^{\phi\phi} \right), \, \delta g^{t\phi}, \, \delta g^{r\theta}  
\label{tadpolesFdeltag2}
\end{equation}
with generic coefficients that can depend on $r$ and $\theta$. As an illustrative example, let us assume that $F$ contains a single factor of $\varepsilon^{tr\theta\phi}$. It is easy to envision how to obtain, for instance,  $\delta g^{t\phi}$ by contracting the $r$ and $\theta$ indices in  $\varepsilon^{tr\theta\phi}$ with an $r$ and a $\theta$-derivative (recall that every index of each type must appear an even number of times). On the other hand, it is clearly not possible to generate  $\delta g^{tr}$ in \eqref{tadpolesFdeltag2} at linear order in $a$, given the ansatz  \eqref{eq:background}.
As a result, the only tadpoles are the ones shown in \eqref{tadpolesFdeltag2}. Using the fact that the trace of $g_{\mu\nu}$ is a constant, one can conveniently trade one of tadpoles in \eqref{tadpolesFdeltag2} for a generic  function of $r$ and $\theta$ at the zero-th order in perturbations.  The final result is thus, schematically,
\begin{equation}
\label{eq:tadpoles-3b}
S_{\rm tadpoles} \supseteq \int \D^{4}x \sqrt{- g} \[ \Lambda(r,\theta) + f_1(r,\theta) \delta g^{rr} + f_2(r,\theta) \delta g^{tt} 
+ a \, f_3(r,\theta) \delta g^{t\phi} + a \, f_4(r,\theta) \delta g^{r\theta} 
\] \, ,
\end{equation}
where we emphasized that $\delta g^{t\phi}$ and $\delta g^{r\theta} $ start linearly in $a$. In conclusion, the number of independent operators, up to linear order in perturbations, is $5$. Since one might be mainly interested in studying the perturbations of black holes in theories with the Gauss--Bonnet and/or the Chern--Simons operators, it is convenient to write the linearized action for the perturbations as
\begin{multline}
S =  \int\D^4x \, \sqrt{-g} \Bigg[
 \frac{\Mpl^2}{2} R + \xi(r,\theta)\mathcal{G} + \zeta(r,\theta) \prescript{*}{}{\! R}^{\mu\nu\rho\sigma}R_{\mu\nu\rho\sigma}
\\
- \Lambda(r,\theta) + f_1(r,\theta) \delta g^{rr} + f_2(r,\theta) \delta g^{tt} 
+ a \, f_3(r,\theta) \delta g^{t\phi} + a \, f_4(r,\theta) \delta g^{r\theta} 
\Bigg] \, ,
\label{EFT-ss0-GBCS}
\end{multline}
where $\xi$ and $\zeta $ are the coefficients of the Gauss--Bonnet and
Chern--Simons operators. In general these can be arbitrary functions of $r$ and $\theta$, however in the subclass of shift-symmetric scalar theories (at the leading order in derivatives) each of them is proportional to the scalar field profile $\bar{\Phi}(r,\theta)$, hence they are identical up to a constant factor.

The Lagrangian~\eqref{EFT-ss0-GBCS} already captures many of the  models that have been  previously considered in the literature.
However, at quadratic order in the perturbations, there can clearly be many  other operators affecting the linearized dynamics. Finding a complete set  is a cumbersome, but straightforward, procedure that closely follows our previous derivation in \cite{Franciolini:2018uyq}. We will not derive it here in general, but, in the spirit of what we discussed in the previous sections, we want to  emphasize that additional operators, such as $(\delta g^{rr})^2$ or $\delta g^{rr} \delta K$, might  not be completely irrelevant. In the presence of symmetries and with couplings satisfying the right power counting, they can be as large as the Gauss--Bonnet and the Chern--Simons operators on the considered background. 

Note also that the generic angular dependence in equation~\eqref{EFT-ss0-GBCS}, given the scalar profile \eqref{eq:phibarCS}, generically induces coupling terms between modes corresponding to different orbital numbers $\ell$, in an expansion in spherical harmonics. Differently from the mixing terms already discussed in Section~\ref{sec:quadratic_action}, the additional mixings induced by a non-trivial $\theta$ dependence couple modes with different $\ell$ within the odd (even) sector.

\section{Conclusions}
\label{sec:conclusions}

In this work we have introduced an effective field theory for perturbations of axisymmetric,  slowly rotating spacetimes sourced by a scalar field  coupled to gravity. Although our main focus was on fluctuations around asymptotically flat black holes with scalar hair, our construction is more general and applies, at the linear order in the spin parameter $a$,  to any stationary, slowly rotating, axisymmetric background. For instance, it can be applied to study perturbations of the background describing the exterior of a star (once the boundary conditions at the surface of the object are specified) or even more exotic spacetimes supported by a scalar field~\cite{Franciolini:2018aad}.

To motivate the use of an effective approach for hairy black hole perturbations, we have discussed an explicit model where the Lagrangian contains, in addition to a coupling $\Phi\mathcal{G}$ between the scalar field $\Phi$ and  the Gauss--Bonnet invariant $\mathcal{G}$, a higher derivative cubic galileon operator, $(\partial\Phi)^2\square\Phi$.  Under the   assumption that $(\partial\Phi)^2\square\Phi$ and $\Phi\mathcal{G}$ become strongly coupled at the same scale, we showed that the presence of $(\partial\Phi)^2\square\Phi$ in the theory yields order-one effects at the level of both the background solution and the dynamics of the perturbations, compared to the case where the coupling of this operator is set to zero. Many other operators, beyond our simple toy model, can induce similar effects at the level of the observables. In this sense, our effective theory is a convenient model-independent framework that allows to include all these effects in a single shot.
We stress that, in our toy model, the introduction of the cubic galileon interaction is merely a choice.\footnote{One could in principle set its coupling to a value that makes the operator negligible in the dynamics of the black hole, without running into fine tuning problems~\cite{Pirtskhalava:2015nla,Santoni:2018rrx,Noller:2019chl}.} There are indications, however, that the presence of additional operators in the theory beyond just $\Phi\mathcal{G}$ is required by fundamental principles, such as microscopic causality and analyticity of the $S$-matrix \cite{Enrico:inprogress}.

There are a few interesting directions that we did not discuss here and we have left for future research. It would be interesting to use our  effective theory to characterize more systematically the properties of the quasi-normal mode spectrum in theories beyond general relativity, and connect the deviations in the frequencies to the form of the operators in the EFT.
We gave a simple  example in this direction in Section~\ref{sec:recoveriso}, where we showed that certain non-trivial relations among the effective couplings (see eqs.~\eqref{MMtuning}), defining a specific subclass of theories within \eqref{EFT-ss0-GB}, allows to recover a degeneracy between the even and odd spectra, which   is in general otherwise broken \cite{Blazquez-Salcedo:2016enn,Bryant:2021xdh}.
Another possibility is along the lines of what we discussed in~\cite{Franciolini:2018uyq}.  There, assuming small deviations from general relativity and introducing a light-ring expansion, we used  the WKB formula of~\cite{Schutz:1985km,Iyer:1986np} to infer some properties of the effective parameters from the   spectrum. Our analysis applied, however, only to the case of a single Schr\"odinger-like equation. The result does not straightforwardly extend to a system of coupled equations, like the one we have in the even sector, or in theories with parity-breaking  operators (see Section~\ref{sec:EFTCS} above). We will come back to this in a separate work \cite{uspaper2}.\footnote{A different way of characterizing deviations from general relativity in the QNM spectrum  was proposed in~\cite{Cardoso:2019mqo,McManus:2019ulj}.  In those works, the field potential in the  Schr\"odinger-like equation  is expanded as a power series  in the radial coordinate, with small coefficients that embody generic deviations in both the background and  the dynamics of the perturbations with respect to general relativity, without making  any reference to an explicit Lagrangian.}

Another important open question is about the generalization of the effective theory to rotating spacetimes, beyond the linear order in $a$. This would require revisiting our construction,  identifying the new complete set of operators for the metric perturbations around the chosen ansatz for the  rotating background metric (see, e.g.,~\cite{Johannsen:2013szh,Carson:2020dez}) and  understanding how to extend the Newman--Penrose formalism in the presence of an extra scalar field that couples to the graviton degrees of freedom to extract the linearized equations of motion.

It would also be interesting to apply our effective theory to more exotic spacetimes and check for instance how the conclusions of~\cite{Franciolini:2018aad} (see also~\cite{Mironov:2018uou}) would get modified for a rotating background.\footnote{Note that, for symmetry reasons, at the linear order in $a$ the equation for the $\ell=0$ mode is unchanged with respect to the non-rotating case ($a=0$), and therefore the conclusions of~\cite{Franciolini:2018aad} would  extend straightforwardly. This could change though for generic rotation, at nonlinear order in $a$.}

\paragraph{Acknowledgements.}  ET thanks Giovanni Maria Tomaselli for discussions. LH is supported by the DOE DE-SC0011941 and a Simons Fellowship in Theoretical Physics.
AP is supported by the Simons Foundation Award No.~658906.
LS is supported by the DOE DE-SC0011941 and the Simons Foundation Award No.~555117.  ET is partly supported by the Italian MIUR under contract 2017FMJFMW (PRIN2017).

\appendix

\section{Black holes with galileon hair: odd sector}
\label{app:cubicgalileonhairodd}

The function $\mathcal{N}(r)$ appearing in eq.~\eqref{cgeqQ2} is given by the solution of the following first-order differential equation,
\begin{multline}
\mathcal{N}'(r) = \frac{1}{4 r^3 (r-r_s) \left(r-r_s+ r \tilde{\alpha}^2 (\mathcal{A}+\mathcal{B})\right)}\Bigg[4 r^2 (r-r_s) (r-2 r_s)  \\    +\tilde{\alpha}^2 \Big[ (r-r_s) \Big(2 r_s^2 \big(3 g_3 r^2 (2 r^2-3 r r_s+r_s^2) \varphi'^3+3 g_3 r^3 (r-r_s)^2 \varphi'^2 \varphi''+4 r_s (11 r_s-8 r) \varphi'    \\   +8 r (r-r_s) (r_s \varphi''+r (r_s-r) \varphi''')\big)+r^4 (\mathcal{A}'+ \mathcal{B}')\Big)+r^3 \mathcal{A} (4 r-9 r_s)+r^3 (10 r-9 r_s) \mathcal{B}\Big]\Bigg] \, ,
\end{multline}
where $\varphi$, $\mathcal{A}$ and $\mathcal{B}$ are defined in eqs.~\eqref{barPhi} and \eqref{ABdef}, and  solve eqs.~\eqref{phip} and \eqref{ABansatz}.
The potential $V_Q(r)$ in eq.~\eqref{cgeqQ} is 
\begin{multline}
V_Q = \left(1-\frac{r_s}{r}\right)\left( \frac{\ell(\ell+1)}{r^2} - \frac{3 r_s}{r^3}   \right)
+ \frac{\tilde \alpha^2}{r^6} \Big[ 8 \omega ^2 r^5 r_s^2  (r-r_s) \varphi''
+ g_3 r^3 r_s^2 (r-r_s)^3 \varphi'^2 \varphi''
\\
+ g_3 r^2 r_s^2 (r-r_s) (2 r^2-3 r r_s+r_s^2) \varphi'^3
+4 r_s^2 (r-r_s) \left(2 (\ell^2+\ell-2) r^2-3 (\ell^2+\ell-3) r r_s-3 r_s^2\right) \varphi'
\\
-8 (\ell^2+\ell-2) r^3 r_s^2 (r-r_s) \varphi''+8 (\ell^2+\ell-5) r^2 r_s^3 (r-r_s) \varphi''+r^3 \mathcal{A} (\ell (\ell+1) r-3 r_s) 
\\
+2 r^2 r_s^2 (r-r_s)^2 \left((11 r_s-4 r) \varphi'''+2 r (r-r_s) \varphi^{(4)}\right)
+3 r^3 (r-r_s) \mathcal{B}(r)+30 r r_s^4 (r-r_s) \varphi''
\Big] .
\label{VQoddcg}
\end{multline}
Note that in the limit $\tilde \alpha\rightarrow0$ one recovers the Regge--Wheeler potential \cite{Regge:1957td} of massless odd spin-$2$ modes in general relativity.
In the eikonal limit, $\ell\rightarrow\infty$, the potential \eqref{VQoddcg} considerably simplifies and  reduces to
\begin{multline}
V_Q(\ell\rightarrow\infty) = \left(1-\frac{r_s}{r}\right)\frac{\ell(\ell+1)}{r^2}
+ \frac{\tilde \alpha^2}{r^5} \Big[ 8 \omega ^2 r^4 r_s^2  (r-r_s) \varphi''
\\
+ \ell(\ell+1) \left(
r^3 \mathcal{A}+4 r_s^2 (r-r_s) (2 r-3 r_s) \varphi'-8 r r_s^2 (r-r_s)^2 \varphi''
\right) \Big] + \mathcal{O}(\ell^0) \, ,
\label{VQoddcgeikonal}
\end{multline}
where we kept only terms up to the order $\mathcal{O}(\ell)$.

\section{Radial foliation and geometric decomposition}
\label{app:foliation}

In this appendix, we recall the main ingredients that we used to construct the EFT in the main text. The notation mirrors the one of~\cite{Franciolini:2018uyq,Franciolini:2018aad}. We assume here the $\Phi$ is a scalar, while we refer to Section~\ref{sec:EFTCS} for a discussion about the case in which parity is broken.

The black hole metric breaks spatial translations, while retaining axial symmetry.
Since the scalar $\Phi$ on the background depends only on $r$ at linear order in $a$, it is convenient to define a foliation of the spacetime with $\Phi = \rm constant $ hypersurfaces. 
The unit vector orthogonal to the radial foliation  is given by \cite{Franciolini:2018uyq}
\begin{equation}
n_{\mu} = \dfrac{\nabla_{\mu}\Phi}{\sqrt{\nabla_{\mu}\Phi\nabla^{\mu}\Phi}},
\label{nmu}
\end{equation}
satisfying $n_{\mu}n^{\mu}=1$.
In analogy with the ADM decomposition in general relativity, one can define the induced metric
\begin{equation}
h_{\mu\nu}\equiv g_{\mu\nu}- n_{\mu}n_{\nu}\, .
\end{equation}
In the unitary gauge,  $n_{r}=N$ and $n_{a}=0$, and we can write the metric as:
\begin{equation}
\D s^{2} = N^{2} \D r^{2} + h_{ab} \( \D x^{a}+ N_{a} \D r \) \( \D x^{b} + N^{b} \D r \),
\end{equation}
where the Latin indices $a,b,c$ denote temporal and angular coordinates, while $N$ and $N^{a}$ are the \emph{lapse} and the \emph{shift}, respectively.
Equivalently,
\begin{equation}
g_{\mu\nu} = 
\begin{pmatrix}
h_{ab} & N_{a}\\
N_{b} & N^{2}+N^{c}N_{c}
\end{pmatrix},
\qquad 
h_{\mu\nu} = 
\begin{pmatrix}
h_{ab} & N_{a}\\
N_{b} & N^{c}N_{c}
\end{pmatrix}.
\end{equation}
By the definitions of $h_{\mu\nu}$ and $n_{\mu}$ the following orthogonality conditions follow:
\begin{equation}
\label{eq:ort_n}
h^{\mu}_{\nu}n_{\mu}=0, \qquad n^{\nu}\nabla_{\mu}n_{\nu}=0\, .
\end{equation}
The extrinsic curvature is given by:
\begin{equation}
\label{eq:ext_curv}
K_{\mu\nu} = h^{\alpha}_{\mu}h^{\beta}_{\nu} \nabla_{\alpha}n_{\beta} = h^{\alpha}_{\mu}\nabla_{\alpha}n_{\nu}=\nabla_{\mu}n_{\nu}-n^{\alpha }n_{\mu}\nabla_{\alpha}n_{\nu} \, .
\end{equation}
Two orthogonality conditions  for $K_{\mu\nu}$ follow from~\eqref{eq:ort_n} and the definition~\eqref{eq:ext_curv}, 
\begin{equation}
\label{eq:ort_k}
n^{\mu}K_{\mu\nu}=n^{\nu}K_{\mu\nu}=0\, .
\end{equation}
The temporal and angular components of the extrinsic curvature can be expressed as
\begin{equation}
K_{ab}=\nabla_{a}n_{b}=-N\Gamma_{ab}^{r} = \dfrac{1}{2N}\(\partial_{r}h_{ab}-D_{a}N_{b}-D_{b}N_{a}\) \, ,
\end{equation}
which is manifestly symmetric in $(a,b)$ and where $D_{a}$ is the covariant derivative acting on the $\(2+1\)$-dimensional hypersurface:
\begin{equation}
D_{a}V_{b}=h^{\mu}_{a}h^{\nu}_{b}\nabla_{\mu}V_{\nu}\, .
\end{equation}
From equation~\eqref{eq:ext_curv} it follows that the trace of the extrinsic curvature can be expressed as a total derivative:
\begin{equation}
K =\nabla_{\mu}n^{\mu}.
\end{equation}

\subsection{Gauss--Codazzi equation}
The $(2+1)$-dimensional Riemann tensor,
\begin{equation}
- \hat{R}_{\alpha\beta\mu\nu}V^{\alpha}=D_{\mu}D_{\nu}V_{\beta}-D_{\nu}D_{\mu}V_{\beta} \, ,
\end{equation}
can be related to the four-dimensional Riemann tensor and the extrinsic curvature through the Gauss--Codazzi equation
\begin{equation}
h^\tau_\mu h^\rho_\nu h^\sigma_\beta R_{\alpha\sigma\tau\rho}
= \hat{R}_{\alpha\beta\mu\nu}  + K_{\mu\beta}K_{\nu\alpha} - K_{\nu\beta} K_{\mu\alpha},
\end{equation}
and the contracted form
\begin{equation}
R = \hat{R} - K_{\mu\nu}K^{\mu\nu} + K^{2} - 2\nabla_{\mu}\(K n^{\mu}- n^{\nu}\nabla_{\nu}n^{\mu} \) \, .
\end{equation}
\subsection{Geometric quantities evaluated on the background}

On the background~\eqref{eq:background} we are considering there are no $(r,a)$ components and many simplifications occur. The shift are zero $\bar{N}_{a}=0$, while the lapse is $\bar{N}=1/\sqrt{B}$.

The induced metric takes the simple form
\begin{equation}
\bar{h}_{\mu\nu} = 
\begin{pmatrix}
\bar{g}_{ab} & 0\\
0 & 0
\end{pmatrix},
\end{equation}
while the temporal and angular components of the extrinsic curvature are given by
\begin{equation}
\bar{K}_{ab}= \dfrac{\sqrt{B}}{2} \partial_{r}\bar{h}_{ab} = \dfrac{\sqrt{B}}{2} \partial_{r}\bar{g}_{ab} 
\end{equation}
We provide now some components of the Christoffel symbols evaluated on the background that will be useful in what follows. At zeroth order in $a$ we have:
\begin{equation}
\label{eq:christoffel_background}
\begin{split}
\bar{\Gamma}_{t\phi}^{\alpha} &= \dfrac{1}{2} \bar{g}^{\alpha\sigma} \( \partial_{t} \bar{g}_{\phi\sigma}+ \partial_{\phi} \bar{g}_{t\sigma}- \partial_{\sigma} \bar{g}_{t\phi} \) = 0 \, , \\
\bar{\Gamma}_{t r}^{\alpha} &= \dfrac{1}{2} \bar{g}^{\alpha\sigma} \( \partial_{t} \bar{g}_{r\sigma}+ \partial_{r} \bar{g}_{t\sigma}- \partial_{\sigma} \bar{g}_{t r} \) = 0 \, , \qquad {\rm for} \; \alpha\neq t \, , \\
\bar{\Gamma}_{t r}^{t} &= \dfrac{1}{2} \bar{g}^{tt}  \partial_{r} \bar{g}_{tt} \, , \\
\bar{\Gamma}_{\phi r}^{\alpha} &= \dfrac{1}{2} \bar{g}^{\alpha\sigma} \( \partial_{\phi} \bar{g}_{r\sigma}+ \partial_{r} \bar{g}_{\phi\sigma}- \partial_{\sigma} \bar{g}_{\phi r} \) = 0 \, , \qquad {\rm for} \; \alpha\neq \phi \, , \\
\bar{\Gamma}_{\phi r}^{\phi} &= \dfrac{1}{2} \bar{g}^{\phi\phi}  \partial_{r} \bar{g}_{\phi\phi}  \, ,
\end{split}
\end{equation}
where we used that $\partial_{t}\bar{g}_{\mu\nu}=\partial_{\phi}\bar{g}_{\mu\nu}=0$, $\bar{g}_{r a}=0$ and $\bar{g}_{t\phi}= \mathcal{O}(a)$.

\section{Infinitesimal variations}
\label{app:infvar}
We collect here results on the infinitesimal variation of various tensors under an infinitesimal transformation $g_{\mu\nu} \rightarrow g_{\mu\nu} + \delta g_{\mu\nu}$. In this section all the terms without a $\delta$ in front are intended to be evaluated on the background.
From $g_{\mu\alpha}g^{\alpha\nu}=\delta_{\mu}^{\nu}$ follows
\begin{equation}
 g^{\mu\nu}\delta g_{\mu\nu} = -g_{\mu\nu} \delta g^{\mu\nu}.
 \end{equation} 
Moreover, from the relation for the determinant of an operator, $\log (\det M ) = {\rm Tr} ( \log M )$, it follows
\begin{equation}
\delta g = g\,g^{\mu\nu}\delta g_{\mu\nu} = - g\, g_{\mu\nu} \delta g^{\mu\nu}.
\end{equation}
The variation of the volume element is easily computed as 
\begin{equation}
\delta \sqrt{-g} = \frac{1}{2} \sqrt{-g} \, g^{\mu\nu}\delta g_{\mu\nu} = -\frac{1}{2} \sqrt{-g} \, g_{\mu\nu} \delta g^{\mu\nu}.
\end{equation}
For the Christoffel symbol we obtain:
\begin{equation}
\begin{split}
\delta \Gamma^{\rho}_{\mu\nu} &=
\dfrac{1}{2} g^{\rho\sigma}\(\nabla_{\mu}\delta g_{\nu\sigma} + \nabla_{\nu} \delta g_{\mu\sigma} - \nabla_{\sigma}\delta g_{\mu\nu} \)\\
&= - \dfrac{1}{2} g^{\rho\sigma}\(g_{\alpha\nu}g_{\beta\sigma}\nabla_{\mu}+g_{\alpha\mu}g_{\beta\sigma}\nabla_{\nu}-g_{\alpha\mu}g_{\beta\nu}\nabla_{\sigma}\) \delta g^{\alpha\beta},
\end{split}
\end{equation}
which implies the variation of the Riemann tensor is
\begin{equation}
\label{eq:riemann_variation}
\delta R_{\mu\rho\nu\sigma} = - R_{\beta\rho\nu\sigma} g_{\alpha\mu} \delta g^{\alpha\beta} + g_{\mu\lambda} \(\nabla_{\nu}\delta \Gamma^{\lambda}_{\rho\sigma}-\nabla_{\sigma}\delta \Gamma^{\lambda}_{\rho\nu}\).
\end{equation}
The variation of the Ricci tensor and its contracted version are:
\begin{equation}
\delta R_{\mu\nu} = \nabla_{\rho} \( \delta\Gamma_{\nu\mu}^{\rho} \) -  \nabla_{\nu} \( \delta\Gamma_{\rho\mu}^{\rho} \),
\end{equation}
\begin{equation}
\begin{split}
g^{\mu\nu}\delta R_{\mu\nu} &= \nabla^{\mu}\nabla^{\nu}(\delta g_{\mu\nu})-\nabla^{\mu}g^{\alpha\beta}\nabla_{\mu}(\delta g_{\alpha\beta}) = \nabla^{\mu}v_{\mu},\\
&= \( -\nabla_{\mu}\nabla_{\nu} + g_{\mu\nu}\Box \) \delta g^{\mu\nu}
\end{split}
\end{equation}
which usually gives rise to boundary terms  that can be neglected in a spacetime without boundaries.

For any scalar of the form
\begin{equation}
X = \nabla_{\mu}X^{\mu}= \dfrac{1}{\sqrt{-g}}\partial_{\mu}\(\sqrt{-g}X^{\mu}\),
\end{equation}
the variation is
\begin{equation}
\delta X = \dfrac{1}{2} g_{\mu\nu}\delta g^{\mu\nu} X +  \nabla_{\mu}\[\delta X^{\mu} - \dfrac{1}{2}X^{\mu} g_{\alpha\beta}\delta g^{\alpha\beta} \].
\end{equation}
For the foliation one has:
\begin{equation}
\delta n_{\mu} = -\dfrac{1}{2} n_{\mu}n_{\alpha}n_{\beta}\delta g^{\alpha\beta},
\end{equation}
\begin{equation}
\delta n^{\mu} = n_{\nu}\delta^{\mu\nu} -\dfrac{1}{2} n^{\mu}n_{\alpha}n_{\beta}\delta g^{\alpha\beta},
\end{equation}
\begin{equation}
\begin{split}
\delta K_{\mu\nu} = &-\frac{1}{2}\nabla_\mu \left(n_\nu n_\alpha n_\beta\delta g^{\alpha\beta}\right) \\
&+ \frac{1}{2}\left(g_{\alpha\nu}n_\beta\nabla_\mu
+ g_{\alpha\mu}n_\beta\nabla_\nu- g_{\alpha\mu}g_{\beta\nu}n^{\rho}\nabla_\rho \right)\delta g^{\alpha\beta} \\
&- n_\beta n_\mu(\nabla_\alpha n_\nu)\delta g^{\alpha\beta} + n_\alpha n_\beta n_\mu n^\rho (\nabla_\rho n_\nu)\delta g^{\alpha\beta}\\
&+ \frac{1}{2} n_\mu n^\rho \nabla_\rho \left( n_\nu n_\alpha n_\beta \delta g^{\alpha\beta} \right)
- \frac{1}{2} n_\alpha n_\beta n_\mu\nabla_\nu\delta g^{\alpha\beta} \, ,
\end{split}
\end{equation}
\begin{equation}
\delta K 
	= \frac{1}{2} g_{\mu\nu}\delta g^{\mu\nu} K + \nabla_\mu\[ 
	 n_\nu\delta g^{\mu\nu} -\frac{1}{2}n^\mu(n_\alpha n_\beta +g_{\alpha\beta})\delta g^{\alpha\beta} 
	 \],
\end{equation}
\begin{equation}
\delta h^{\mu\nu} = \delta g^{\mu\nu} - n_\lambda\( n^\mu\delta g^{\nu\lambda} + n^\nu\delta g^{\mu\lambda} \) + n^\mu n^\nu n_\alpha n_\beta\delta g^{\alpha\beta}
\end{equation}
\begin{equation}
\begin{split}
\delta\hat{R}
=&  \left( R_{\alpha\beta} - n^\mu n^\nu R_{\mu\alpha\nu\beta}
	- 3R_{\rho\alpha} n_\beta n^\rho
	+ 2R_{\mu\nu} n^\mu n^\nu n_\alpha n_\beta
	 \right) \delta g^{\alpha\beta}\\
&+ h^{\nu}_\mu h^{\rho\sigma} \( \nabla_\nu\delta\Gamma^\mu_{\rho\sigma}  - \nabla_\sigma\delta\Gamma^\mu_{\rho\nu} 
\)
	+ 2K\delta K - 2K^{\mu\nu}\delta K_{\mu\nu}.
\end{split}
\end{equation}
The following identity is satisfied:
\begin{equation}
n_{\mu}\delta h^{\mu\nu}=h^{\mu\nu}\delta n_{\mu}=0.
\end{equation}
We provide an explicit expression for the variation of a component of the Christoffel symbol which will be of use in what follows:
\begin{equation}
\label{eq:christ_var1}
\begin{split}
\delta \Gamma^{r}_{t\phi} &= \dfrac{1}{2} g^{r\sigma}\(\nabla_{t}\delta g_{\phi\sigma} + \nabla_{\phi} \delta g_{t\sigma} - \nabla_{\sigma}\delta g_{t\phi} \)\\
&= \dfrac{1}{2} g^{rr}\(\nabla_{t}\delta g_{\phi r} + \nabla_{\phi} \delta g_{tr} - \nabla_{t}\delta g_{t\phi} \),
\end{split}
\end{equation}
where we used $\bar{g}_{r\sigma}= 0$ for $\sigma\neq r$. At linear order in $a$, using the results of equation~\eqref{eq:christoffel_background} we obtain:
\begin{equation}
\label{eq:christ_var2}
\begin{split}
\nabla_{t} \delta g_{\phi r} &= \partial_{t} \delta g_{\phi r} - \bar{\Gamma}_{t\phi}^{\alpha} \delta g_{\alpha r} - \bar{\Gamma}_{tr}^{\alpha}\delta g_{\phi\beta}\\
&=\partial_{t} \delta g_{\phi r} - \bar{\Gamma}_{tr}^{t}\delta g_{\phi t} \, ,\\
\nabla_{\phi} \delta g_{t r} &= \partial_{\phi} \delta g_{t r} - \bar{\Gamma}_{\phi t}^{\alpha} \delta g_{\alpha r} - \bar{\Gamma}_{\phi r}^{\alpha}\delta g_{t\beta}\\
&=\partial_{\phi} \delta g_{t r} - \bar{\Gamma}_{\phi r}^{\phi}\delta g_{t \phi} \, ,\\
\nabla_{r} \delta g_{t\phi } &= \partial_{r} \delta g_{t\phi } - \bar{\Gamma}_{tr}^{\alpha} \delta g_{\alpha \phi} - \bar{\Gamma}_{t\phi}^{\alpha}\delta g_{r\beta} \\
&=\partial_{r} \delta g_{t\phi} - \bar{\Gamma}_{tr}^{t}\delta g_{t\phi} \, .
\end{split}
\end{equation}

Another useful expression is the variation of the  Gauss--Bonnet operator:
\begin{equation}
\begin{split}
\int   \D^4 x\sqrt{-g} \, \xi(r) \delta \mathcal{G} 
= \int \D^4 x  & \sqrt{-g} \, \Big[
  2 \xi (R R_{\alpha\beta} - {R_\alpha}^{\rho\nu\sigma} R_{\beta\rho\nu\sigma} ) + 4 \nabla_\rho\nabla_\sigma(\xi {{{R_\alpha}^\rho}_\beta}^\sigma)
\\
& + 8 \nabla_\lambda \nabla_\beta (\xi {R_\alpha}^\lambda)
- 2 \nabla_\alpha\nabla_\beta (\xi R)
-4 \nabla_\lambda\nabla^\lambda (\xi R_{\alpha\beta})
\\
& - 4 g_{\alpha\beta} \nabla_\mu\nabla_\nu (\xi R^{\mu\nu})
+ 2 g_{\alpha\beta} \nabla_\lambda\nabla^\lambda (\xi R )
\Big] \delta g^{\alpha\beta} \, ,
\end{split}
\end{equation}
where $\xi(r)$ is a generic $r$-dependent function and $\mathcal{G}$ is the Gauss--Bonnet invariant defined in \eqref{GBdefinition}.

\subsection{Equivalent expressions for the $t \phi$ tadpole}
\label{appendix:tphi_tadpole}
We show here that the tadpole $\delta K_{t\phi}$ can be rewritten as  $\delta g^{t\phi}$. By construction, the $t\phi$ indices in $\delta K_{t\phi}$ are implicitly  contracted  with  $\bar{g}_{t\phi}$ or its derivative, which are of order $a$. Therefore, working at linear order in $a$ we can neglect all the terms of order $a$ in the expression for $\delta K_{t\phi}$ as a polynomial in the unit vector, the background metric and their derivatives. By definition, the extrinsic curvature is
\begin{equation}
K_{t\phi}= \nabla_{t}n_{\phi} = \partial_{t} n_{\phi} - \Gamma_{t\phi}^{\alpha}n_{\alpha}.
\end{equation}
Considering its perturbation, at zeroth order in $a$ we have:
\begin{equation}
\label{eq:KGamma_tphi}
\delta K_{t\phi} = \partial_{t} \delta n_{\phi}- \delta \Gamma^{\alpha}_{t\phi}\,\bar{n}_{\alpha} - \bar{\Gamma}_{t\phi}^{\alpha}\,\delta n_{\alpha}= - \delta \Gamma^{r}_{t\phi}\,\bar{n}_{r}
\end{equation}
where the first term was eliminated by an integration by parts, using that the background is time independent, the unit vector was set on the background and the last term was set to zero due to eq.~\eqref{eq:christoffel_background}.
Using the results of equations~\eqref{eq:christ_var1} and~\eqref{eq:christ_var2} and getting rid of terms with $t$ and $\phi$ derivatives, by an integration by parts we obtain:
\begin{equation}
\delta K_{t\phi} = f_{1}(r) \partial_{r} \delta g_{t\phi} + f_{2}(r) \delta g_{t\phi}.
\end{equation}
At zeroth order in $a$ one has $\delta g_{t\phi} = \bar{g}_{t\alpha} \bar{g}_{\phi \beta} \delta g^{\alpha\beta} =\bar{g}_{tt} \bar{g}_{\phi\phi}\delta g^{t\phi}$.
Integrating by part the radial derivative we finally arrive at the desired result, i.e., up to boundary terms,
\begin{equation}
\delta K_{t\phi} = f_{3}(r) \delta g^{t\phi}.
\end{equation}

\section{Explicit example}
\label{app:example}

In this appendix we discuss an example of slowly rotating hairy black hole solution in a theory with no higher derivative operators, which can be described at quadratic order by the effective action of Section~\ref{sec:quadratic_action}.
The theory is that of a minimally coupled phantom scalar field with negative kinetic term~\cite{Dennhardt:1996cz}, described by the action
\begin{equation}
S = \int {\rm d}^{4}x \sqrt{-g} \left[ \dfrac{R}{2} + \dfrac{1}{2} g^{\mu\nu} \partial_{\mu}\Phi \partial_{\nu}\Phi - \mathcal{V}(\Phi) \right] \, ,
\end{equation}
where we have set $\Mpl=1$ and where the potential is given by the expression
\begin{equation}
\mathcal{V}(\Phi) = \dfrac{3(q+ 2 \mathcal{M})}{\vert q \vert^{3}} \left[(3+\Phi^{2}) \cdot \sinh \vert \Phi \vert - 3 \vert \Phi \vert \cdot \cosh \Phi \right].
\end{equation}
Even if not well motivated from an EFT point of view due to its ghost-like kinetic term and the specific form of the potential, this theory serves the purpose of providing a simple example of hairy black hole which can be treated analytically. The spherically symmetric hairy black hole solution depends on two parameters, the asymptotic mass $8\pi \mathcal{M}$ and the scalar charge $q$, and was found in Ref.~\cite{Dennhardt:1996cz} to be: 
\begin{equation}
\begin{split}
&\bar{\Phi}(r) = \dfrac{q}{r}, \\
& A(r)= B(r) = \dfrac{r^{2}(6\mathcal{M}+3q) \e^{-\frac{q}{r}}}{4q^{3}}-\left[\dfrac{r^{2}(6\mathcal{M}+3q)}{4q^{3}} - \dfrac{r(6\mathcal{M}+3q)}{2q^{2}} +\dfrac{6\mathcal{M}+q}{2q} \right] \e^{\frac{q}{r}}, \\
& C(r) = r^{2} \e^{-\frac{q}{r}}.
\end{split}
\end{equation}
This solution can be straightforwardly generalized to a slowly rotating hairy black hole. Using the background equation~\eqref{eq:tadpoleD} derived in Section~\ref{sec:tadpoles}, we find that at first order in $a$ the functions $\bar{\Phi}(r),  A(r), B(r),C(r)$ are unmodified, while
\begin{equation}
D(r) = \dfrac{6\mathcal{M}}{4q^{3}} \left[ (2q^{2}-2 q r + r^{2}) \e^{\frac{q}{r}}- r^{2} \e^{-\frac{q}{r}} \right].
\end{equation}
This solution reduces to a slowly rotating Kerr black hole in the limit $q\rightarrow 0$.

Another closed form example of slowly rotating hairy black hole has been recently presented in Ref.~\cite{Bakopoulos:2021dry}, generalizing a solution first found in Ref.~\cite{Herdeiro:2015waa} in a theory with a scalar field minimally coupled to gravity and with no higher derivate operators. The no-hair theorem of Bekenstein~\cite{Bekenstein:1995un} is evaded thanks to a negative potential, $\mathcal{V}(\Phi)<0$, which in general corresponds to a violation of the weak energy condition. We refer the reader to Ref.~\cite{Bakopoulos:2021dry} for the explicit form of the solution.

As expected, the background metric and the scalar hair profile in both these examples respect the form of our ansatz~\eqref{eq:background}. Even though they are not particularly physical, being based on theories with undesirable properties, such as a negative kinetic term or potential, they can be thought of as illustrative examples that show how the EFT works in the simplest possible case. Indeed, the dynamics of these theories can be captured in a unified way by our EFT, and in particular by its simplest form, corresponding to the leading order in the derivative expansion detailed in Section~\ref{sec:quadratic_action}.

\section{Vector and tensor spherical harmonics, and some useful identities}
\label{app:spherical_harmonics}

In problems with spherical symmetry it is convenient to express quantities in terms of spherical coordinates and expand the angular dependence in terms of spherical harmonics. In the case of slowly rotating backgrounds, however, the spherical symmetry is weakly broken by the spin parameter $a$. Spherical harmonics can still be a useful basis of functions to parametrize the angular dependence, but symmetry breaking effects can in general induce mixing between different harmonics. Because of this it is useful to work in terms of an orthonormal basis of functions, so that it is straightforward to project on the relevant basis components when needed.

We briefly review the construction of vector and tensor spherical harmonics, following~\cite{Regge:1957td}, to fix our normalizations (see also Appendix A of Ref.~\cite{Hui:2020xxx}). Let us consider the 2-sphere $S^{2}$, parametrized by the coordinates $\theta,\phi$, with $\theta\in [0,\pi]$ and $\phi \in [0,2\pi)$. The Latin indices $i,j,\dots$ will run on $\theta,\phi$. We will use the standard metric and Levi--Civita tensors on the sphere: 
\begin{equation}
\gamma_{ij}=
\begin{pmatrix}
1 & 0\\
0 &\sin^{2}\theta
\end{pmatrix}, \qquad
\varepsilon_{ij}=
\sin\theta \begin{pmatrix}
0 & +1\\
-1 & 0
\end{pmatrix}.
\label{gijLCt}
\end{equation}
We denote the scalar spherical harmonics as $Y^{(\ell,m)}(\theta,\phi)$, defined as 
\begin{equation}
Y^{(\ell,m)}(\theta,\phi) = (-1)^{m} \sqrt{\dfrac{2 \ell +1}{4\pi}} \sqrt{\dfrac{(\ell -m)!}{(\ell + m)!}} \; P^{(\ell,m)}(\cos \theta) \e^{i m \phi},
\end{equation}
where the associated Legendre functions are given by Rodrigues' formula:
\begin{equation}
P^{(\ell,m)}(x) = \dfrac{1}{2^{\ell} \ell !} (1-x^{2})^{\frac{m}{2}} \dfrac{{\rm d}^{\ell +m}}{{\rm d} x^{\ell +m}}(x^{2}-1)^{\ell}
\end{equation}
and satisfy the associated Legendre equation.
The spherical harmonics so defined are orthonormal when integrated on the solid angle with ${\rm d}\Omega = \sin\theta {\rm d}\theta{\rm d}\phi$:
\begin{equation}
\int {\rm d}\Omega \; Y^{(\ell,m)}(\theta,\phi)^* \;Y^{(\ell',m')}(\theta,\phi) = \delta_{\ell\ell' } \delta_{m m' }.
\end{equation} 
An orthonormal basis of vector and tensor spherical harmonics is given by:
\begin{equation}
\begin{split}
& ^{(+)}Y^{(\ell,m)}_{i} = \dfrac{1}{\sqrt{\ell(\ell +1)}} \; \nabla_{i} Y^{(\ell,m)}, \\
& ^{(-)}Y^{(\ell,m)}_{i} = \dfrac{1}{\sqrt{\ell(\ell +1)}} \;\varepsilon_{i}^{\; j}\nabla_{j} Y^{(\ell,m)}, \\
& ^{\rm (tr)}Y^{(\ell,m)}_{ij} = \dfrac{1}{\sqrt{2}} \; \gamma_{ij} Y^{(\ell,m)}, \\
& ^{\rm (+)}Y^{(\ell,m)}_{ij} = \dfrac{\sqrt{2}}{\sqrt{\ell(\ell+1)(\ell^{2}+\ell -2)}} \; \bigg(\nabla_{i}\nabla_{j} Y^{(\ell,m)} -\dfrac{1}{2} \gamma_{ij} \nabla^{k}\nabla_{k} Y^{(\ell,m)}\bigg), \\
& ^{\rm (-)}Y^{(\ell,m)}_{ij} = \dfrac{1}{2} \bigg(\varepsilon_{i}^{\; k} \;{}^{\rm (+)}Y^{(\ell,m)}_{kj} + \varepsilon_{j}^{\; k}\; {}^{\rm (+)}Y^{(\ell,m)}_{ki} \bigg),
\end{split}
\end{equation}
with scalar product corresponding to a trace on the discrete variables and an integral on the solid angle for the continuous ones.

\subsection{Recurrence relations and other useful identities}

The (associated) Legendre functions and the spherical harmonics satisfy a rich set of recurrence relations and identities, see for instance~\cite{ARFKEN2013715}. When dealing with problems with spherical symmetry it is usually sufficient to consider the simpler identities satisfied by Legendre polynomials, corresponding (up to factors) to spherical harmonics with $m=0$. In the case of a (slowly) rotating background, spherical symmetry is broken explicitly and we will need more general identities on spherical harmonics in order to explicitly express angular functions as combinations of (scalar, vector or tensor) spherical harmonics.

Two useful recurrence relations can be derived through straightforward manipulations of the relations on associated Legendre functions in~\cite{ARFKEN2013715}:
\begin{align}
&\cos \theta \; Y^{(\ell,m)} = \zeta_{\ell,m} Y^{(\ell -1,m)} + \zeta_{\ell + 1,m} Y^{(\ell +1,m)}, \label{eq:recurrence1a}\\
&\sin \theta \, \dfrac{\partial}{\partial \theta} Y^{(\ell,m)} = - (\ell +1)  \zeta_{\ell,m} Y^{(\ell -1,m)} +\ell \zeta_{\ell + 1,m} Y^{(\ell +1,m)},\label{eq:recurrence1b}
\end{align}
where for notational convenience we introduced the coefficients $$\zeta_{\ell,m}= \sqrt{\dfrac{(\ell + m)(\ell - m)}{(2\ell + 1)(2\ell -1)}}.$$
Other useful identities, relating $Y^{(\ell,m)}$ and its $\theta$-derivative to the $\theta$-derivative of $Y^{(\ell\pm 1,m)}$, can be proved by manipulating the generating function of the associated Legendre functions:
\begin{equation}
g_{m}(x,t) = \dfrac{(2m)! \,(1-x^{2})^{\frac{m}{2}}}{2^{m}(m!) (1- 2x t + t^{2})^{m+\frac{1}{2}}}= \sum_{s=0}^{\infty} P^{(s+m,m)}(x) \, t^{s}.
\end{equation}
In particular, by taking the $x$-derivative of $g_{m}(x,t)$ and combining the resulting recurrence relation with a linear combination of equations~\eqref{eq:recurrence1a} and~\eqref{eq:recurrence1b} and the derivative of equation~\eqref{eq:recurrence1a}, the following identities are obtained:
\begin{align}
\label{eq:recurrence2}
& \sin^{2} \theta \; Y^{(\ell,m)} = \dfrac{m^{2}}{\ell(\ell +1)} Y^{(\ell,m)} + \dfrac{1}{\ell} \zeta_{\ell,m} \sin \theta \, \dfrac{\partial}{\partial \theta} Y^{(\ell -1,m)} -\dfrac{1}{\ell +1} \zeta_{\ell + 1,m} \sin \theta \, \dfrac{\partial}{\partial \theta} Y^{(\ell +1,m)}, \\
& \dfrac{1}{2}\sin 2\theta \, \dfrac{\partial}{\partial \theta} Y^{(\ell,m)} = \dfrac{m^{2}}{\ell(\ell +1)} Y^{(\ell,m)} + \dfrac{\ell+1}{\ell} \zeta_{\ell,m} \sin \theta \, \dfrac{\partial}{\partial \theta} Y^{(\ell -1,m)} +\dfrac{\ell}{\ell+1} \zeta_{\ell + 1,m} \sin \theta \, \dfrac{\partial}{\partial \theta} Y^{(\ell +1,m)}.
\end{align}
The identities reduce to well-known relations on Legendre polynomials $P^{(\ell)}(x)$ for $m=0$.

\section{Gauge choice}
\label{app:gauge}

We decompose a generic metric perturbation around the background~\eqref{eq:background} in terms of vector and tensor harmonics, following~\cite{Regge:1957td}. Using a notation where a Greek index $\mu$ runs on the Boyer--Lindquist coordinates $(t,r,\theta,\phi)$, and suppressing the indices $(\ell,m)$ and the explicit coordinate dependence, but keeping in mind that the scalar functions $\tilde{{\rm h}},\tilde{H},\tilde{\mathcal{H}},\tilde{\mathcal{K}}$ and $\tilde{G}$ depend on $(t,r)$:
\begin{equation}
\setlength{\jot}{10pt} 
\begin{split}
&\delta g^{(\rm odd)}_{\mu\nu}=
\begin{pmatrix}
0 & 0 & \tilde{{\rm h}}_{0}\;^{(-)}Y^{(\ell,m)}_{i}\\
0 & 0 & \tilde{{\rm h}}_{1}\;^{(-)}Y^{(\ell,m)}_{i} \\
({\rm sym}) & ({\rm sym}) & \tilde{{\rm h}}_{2}\; ^{(-)}Y^{(\ell,m)}_{ij}
\end{pmatrix}, \\ 
&\delta g^{(\rm even)}_{\mu\nu}=
\begin{pmatrix}
\tilde{H}_{0}\; Y^{(\ell,m)} & \tilde{H}_{1}\; Y^{(\ell,m)} & \tilde{\mathcal{H}}_{0}\; ^{(+)}Y^{(\ell,m)}_{i}\\
({\rm sym})  & \tilde{H}_{2}\; Y^{(\ell,m)} & \tilde{\mathcal{H}}_{1}\; ^{(+)}Y^{(\ell,m)}_{i} \\
({\rm sym}) & ({\rm sym}) & \tilde{\mathcal{K}}\; ^{(\rm tr)}Y^{(\ell,m)}_{ij} + \tilde{G}\; ^{(+)}Y^{(\ell,m)}_{ij}
\end{pmatrix}.
\end{split}
\end{equation}
Under an infinitesimal diffeomorphism transformation $x_{\mu} \rightarrow x_{\mu} + \,\xi_{\mu}$ the metric perturbation transforms as $\delta g_{\mu\nu} \rightarrow \delta g_{\mu\nu} - \nabla_{\mu} \xi_{\nu} - \nabla_{\nu} \xi_{\mu}$. The displacement vector can be expanded in harmonics as:\footnote{The function $\gamma$ that we use here to parametrize  a gauge transformation of the even type should not be confused with the metric \eqref{gijLCt} in the previous section.}
\begin{equation}
\xi_{\mu}^{(\rm even)} =\left(\alpha \, Y^{(\ell,m)}, \beta\, Y^{(\ell,m)}, \gamma\; ^{(+)}Y^{(\ell,m)}_{i} \right), \qquad \xi_{\mu}^{(\rm odd)} =\left(0, 0, \delta\; ^{(-)}Y^{(\ell,m)}_{i} \right),
\end{equation}
where the functions $\alpha,\beta,\gamma,\delta$ have indices $(\ell,m)$ and depend on $(t,r)$. Since the background breaks explicitly rotational invariance and parity, mixing between parity-even and parity-odd terms and mixing between different harmonics with angular number $\ell$ will be generated. For notational convenience we reintroduce the index $\ell$ to account for the mixing, but suppress the index $m$. All the metric functions and gauge parameters are understood to be functions of $(t,r)$, with time derivatives denoted by a dot and radial derivatives denoted by a prime. At linear order in the spin parameter $a$, by using the identities of appendix~\ref{app:spherical_harmonics}, it is straightforward if somewhat lengthy to derive how the metric perturbations transform under an infinitesimal diffeomorphism. We obtain:
\begin{fleqn}[10pt]
\begin{equation}
\begin{split}
&\tilde{{\rm h}}_{0}^{(\ell)} \rightarrow \tilde{{\rm h}}_{0}^{(\ell)} - \dot{\delta}^{(\ell)}\hspace{-5pt}& & - 2 im a \dfrac{1}{\ell(\ell+1)}  D \delta^{(\ell)} \\
& & &+a \dfrac{1}{\ell } \zeta_{\ell ,m} \sqrt{\ell (\ell +1)}  B D' \beta^{(\ell -1)} - a\dfrac{1}{\ell +1} \zeta_{\ell +1,m} \sqrt{\ell (\ell +1)}  B D' \beta^{(\ell +1)} \\
& & &-a \dfrac{2(\ell -1)}{\ell } \zeta_{\ell ,m} \sqrt{\dfrac{\ell +1}{\ell-1}} \dfrac{D}{C} \gamma^{(\ell -1)} - a \dfrac{2(\ell +1)}{\ell +} \zeta_{\ell+1 ,m} \sqrt{\dfrac{\ell}{\ell+2}} \dfrac{D}{C} \gamma^{(\ell +1)},
\end{split}
\end{equation}
\end{fleqn}
\begin{fleqn}[10pt]
\begin{equation}
\begin{split}
&\tilde{{\rm h}}_{1}^{(\ell)} \rightarrow \tilde{{\rm h}}_{1}^{(\ell)} - \delta'{}^{(\ell)} + \dfrac{C'}{C}\delta^{(\ell)} \hspace{-5pt}& & -a \dfrac{1}{\ell } \zeta_{\ell ,m} \sqrt{\ell (\ell +1)}  \dfrac{D C' - C D'}{AC} \alpha^{(\ell -1)} \\
& & &+ a\dfrac{1}{\ell +1} \zeta_{\ell +1,m} \sqrt{\ell (\ell +1)}  \dfrac{D C' - C D'}{AC} \alpha^{(\ell +1)} ,
\end{split}
\end{equation}
\end{fleqn}
\begin{fleqn}[10pt]
\begin{align}
\tilde{{\rm h}}_{2}^{(\ell)} \rightarrow \tilde{{\rm h}}_{2}^{(\ell)} - \sqrt{2} \sqrt{\ell^{2} +\ell -2}\, \delta^{(\ell)}  ,
\end{align}
\end{fleqn}
\begin{fleqn}[10pt]
\begin{align}
\tilde{H}_{0}^{(\ell)} \rightarrow \tilde{H}_{0}^{(\ell)} - 2 \dot{\alpha}^{(\ell)}  +A' B \beta^{(\ell)},
\end{align}
\end{fleqn}
\begin{fleqn}[10pt]
\begin{equation}
\begin{split}
&\tilde{H}_{1}^{(\ell)} \rightarrow \tilde{H}_{1}^{(\ell)} + \dfrac{A'}{A} \alpha^{(\ell)}  -\alpha'{}^{(\ell)}- \dot{\beta}^{(\ell)} \hspace{-5pt}& & - im a \dfrac{1}{\ell(\ell+1)} \dfrac{A D' - A' D}{AC} \gamma^{(\ell)} \\
& & & + a \dfrac{(\ell -1)}{\sqrt{\ell(\ell -1)}} \zeta_{\ell ,m} (A D' - A' D ) \delta^{(\ell -1)}  \\
& & & - a \dfrac{(\ell +2)}{\sqrt{(\ell +1)(\ell +2)}} \zeta_{\ell+1 ,m} (A D' - A' D ) \delta^{(\ell +1)},
\end{split}
\end{equation}
\end{fleqn}
\begin{fleqn}[10pt]
\begin{align}
\tilde{H}_{2}^{(\ell)} \rightarrow \tilde{H}_{2}^{(\ell)} - \dfrac{B'}{B} \beta^{(\ell)} -2  \beta'{}^{(\ell)},
\end{align}
\end{fleqn}
\begin{fleqn}[10pt]
\begin{equation}
\begin{split}
&\tilde{\mathcal{H}}_{0}^{(\ell)} \rightarrow \tilde{\mathcal{H}}_{0}^{(\ell)} - \sqrt{\ell(\ell+1)} \alpha^{(\ell)} - \dot{\gamma}^{(\ell)}\hspace{-5pt}& & - im a \dfrac{1}{\sqrt{\ell(\ell+1)}}  B D' \beta^{(\ell)} - 2 im a \dfrac{1}{\ell(\ell+1)} \dfrac{D}{C} \gamma^{(\ell)} \\
& & &+ 2a \zeta_{\ell ,m} \dfrac{\sqrt{(\ell-1) (\ell +1)}}{\ell } D \delta^{(\ell -1)} \\
& & &+ 2a \zeta_{\ell +1,m} \dfrac{\sqrt{\ell (\ell +2)}}{\ell +1} D \delta^{(\ell +1)} ,
\end{split}
\end{equation}
\end{fleqn}
\begin{fleqn}[10pt]
\begin{align}
\tilde{\mathcal{H}}_{1}^{(\ell)} \rightarrow \tilde{\mathcal{H}}_{1}^{(\ell)} - \sqrt{\ell(\ell+1)} \beta^{(\ell)} - \gamma'{}^{(\ell)}+ \dfrac{C'}{C} \gamma^{(\ell)}\hspace{5pt} + im a \dfrac{1}{\sqrt{\ell(\ell+1)}}  \dfrac{DC'-CD'}{AC} \alpha^{(\ell)} ,
\end{align}
\end{fleqn}
\begin{fleqn}[10pt]
\begin{align}
\tilde{\mathcal{K}}^{(\ell)} \rightarrow \tilde{\mathcal{K}}^{(\ell)} +\sqrt{2}\sqrt{\ell(\ell+1)} \gamma^{(\ell)} - \sqrt{2} B C' \beta^{(\ell)},
\end{align}
\end{fleqn}
\begin{fleqn}[10pt]
\begin{align}
\tilde{G}^{(\ell)} \rightarrow \tilde{G}^{(\ell)} -\sqrt{2} \sqrt{\ell^{2} +\ell -2} \gamma^{(\ell)}.
\end{align}
\end{fleqn}
As expected on symmetry grounds, at linear order in $a$ perturbations of a given parity and fixed $\ell$ can receive contributions from opposite parity gauge transformations with $\ell \pm 1$ and from the $\phi$-derivative of same parity gauge transformations.
Moreover, at linear order in $a$ and neglecting parity-breaking contributions for simplicity, the scalar field perturbation transforms as 
\begin{equation}
\delta\Phi^{(\ell)} \rightarrow \delta \Phi^{(\ell)} + \beta^{(\ell)}  \bar{\Phi}',
\end{equation}
where we have again expanded a scalar perturbation in spherical harmonics and suppressed the $m$ index.

The notation used in this appendix is related to the notation used in the rest of the article (chosen to agree with that of Ref.~\cite{Franciolini:2018uyq}) as follows:
\begin{equation}
\begin{split}
&{\rm h}_{0}= \sqrt{\ell (\ell+1)} \tilde{{\rm h}}_{0}, \qquad {\rm h}_{1}= \sqrt{\ell (\ell+1)} \tilde{{\rm h}}_{1}, \qquad {\rm h}_{2}= \dfrac{\sqrt{2}}{\sqrt{\ell (\ell+1)(\ell^{2}+\ell -2)}} \tilde{{\rm h}}_{2}, \\
&H_{0}= \tilde{H}_{0}/A, \quad \qquad \qquad H_{1}= \tilde{H}_{1}, \quad \qquad \qquad H_{2}= B \tilde{H}_{2}, \\
&\mathcal{H}_{0}= \sqrt{\ell (\ell+1)} \tilde{\mathcal{H}}_{0}, \qquad \mathcal{H}_{1}= \sqrt{\ell (\ell+1)} \tilde{\mathcal{H}}_{1}, \\
& G = \dfrac{\sqrt{2}}{C\sqrt{\ell (\ell+1)(\ell^{2}+\ell -2)}} \tilde{G}, \qquad \mathcal{K} = - \dfrac{\sqrt{\ell (\ell+1)}}{C\sqrt{2(\ell^{2}+\ell -2)}} \tilde{G} + \dfrac{1}{C \sqrt{2}} \tilde{\mathcal{K}}.
\end{split}
\end{equation}
We choose to work in the Regge--Wheeler-unitary gauge, fixing
\begin{equation}
\delta \Phi =\mathcal{H}_0 = G = {\rm h}_2=0.
\end{equation}
The conditions on $\delta \Phi$, $G$, ${\rm h}_2$ determine in a unique way $\beta^{(\ell)}$, $\gamma^{(\ell)}$, $\delta^{(\ell)}$ respectively. The condition on $\mathcal{H}_0$ fixes $\alpha^{(\ell)}$ in terms of $\beta^{(\ell)}$, $\gamma^{(\ell)}$, $\delta^{(\ell \pm 1)}$.
Since the gauge fixing is complete---i.e., it determines the gauge parameters in a unique way without ambiguity---it is consistent to fix the gauge at the level of the action without losing any constraint, see for instance Ref.~\cite{Motohashi:2016prk}.

\section{Linearized equations of motion with even-odd mixing}
\label{app:eoms}

In this appendix we derive the equations of motion for the theory \eqref{EFT0der}, where we retained   only the operators at the leading order in the derivative expansion. Similar manipulations hold in the presence of other quadratic operators in the EFT.

\paragraph{Modes with $\ell\geq2$.} Let us start assuming $\ell$ generic and  $\ell\geq2$. 
We start from the parametrization \eqref{metricperts}, where $\delta g_{\mu\nu}^{\rm odd}$ and  $\delta g_{\mu\nu}^{\rm even}$ are given in \eqref{odd metric perturbations version 0} and  \eqref{even metric perturbations version 0}. Then, we shall fix the gauge where $\mathcal{H}_0=G={\rm h}_2=0$ and expand the action up to quadratic order in the fields.\footnote{As discussed in appendix~\ref{app:gauge}, it is consistent to fix the  gauge $\mathcal{H}_0=G={\rm h}_2=0$ directly at the level of the action \eqref{EFT0der}. }
From the quadratic action, we can compute the equations of motion for the remaining fields $H_0$, $H_1$, $H_2$, $\mathcal{H}_1$, $\mathcal{K}$, ${\rm h}_0$ and ${\rm h}_1$.
To integrate out more easily the constraint variables, it is convenient to redefine the metric perturbation $H_2$ as follows~\cite{Kobayashi:2012kh,Kobayashi:2014wsa,Franciolini:2018uyq,Franciolini:2018aad},
\begin{equation}
H_2 = \psi + \frac{2 \ell (\ell+1) }{B(r) C'(r)} \mathcal{H}_1  +\frac{2 C(r) }{B(r) C'(r)}\mathcal{K}'  \, .
\label{H2fr}
\end{equation}
This allows to get rid of  $\mathcal{H}_1' $ and $\mathcal{K}''$ from the $H_0$'s equation of motion. The resulting  set of coupled equations is:
\begin{multline}
0=
Y_{\ell m}\Bigg[
\mathcal{H}_1 \left(2 \ell (\ell+1) \left( C'  -\frac{ C A'}{A}  \right)
-\frac{4 \ell (\ell+1)^2 C}{BC'}
  +  2 a m \omega \left(   \frac{  DC'}{A}-\frac{ CD'}{A}   \right)\right)
\\
+\mathcal{K}' \left(-\frac{2C^2 A'}{A}-\frac{4 \ell (\ell+1) C^2}{BC'}+2CC'\right)
+\mathcal{K} \left(\frac{4 C}{B}-\frac{2 \ell (\ell+1)C}{B}\right)
\\
-2 BCC' \psi '
+ \psi  \left(-\frac{BC A'C'}{A}- 3 B'C'C  - 2 \ell (\ell+1)C -2 BCC''\right)
\\
+ 2 i a m H_1 \left(\frac{ DC'}{A}-\frac{ C D'}{A}\right)
\Bigg]
+  
2 a \sin\theta (\partial_\theta Y_{\ell m})\Bigg[
{\rm h'_0} \left(\frac{DC'}{A}-\frac{CD'}{A}\right)
\\
+{\rm h_0} \left(\frac{C'D'}{A}-\frac{DC'^2}{AC}\right)
+ i \omega {\rm h_1} \left(\frac{  CD'}{A}-\frac{ DC'}{A}\right)
\Bigg] \, ,
\end{multline}
\begin{multline}
0=
Y_{\ell m}\Bigg[
i a m \mathcal{K}'\frac{ CD'}{2 AC'}
+i \omega \mathcal{K} \left(\frac{  C A'}{2 A^2}-\frac{ C'}{2 A}\right)
+\psi (r) \left(\frac{i a m BDC'}{4 AC}+\frac{i a m BD'}{4 A}+\frac{i \omega  BC'}{2 A}\right)
\\
+    \mathcal{H}_1 \ell(\ell+1) \left(  i a  m\left( \frac{D'}{2 AC'}+\frac{D}{2 AC}
\right)
+\frac{i  \omega}{2 A}\right)
+ H_0 \left(\frac{i a mD'}{4 A}-\frac{i a m DC'}{4 AC}\right)
\\
+H_1 \frac{\ell(\ell+1)}{2 A}
+a \,  {\rm h_1} \cos\theta \frac{\ell(\ell+1)  D}{AC}
\Bigg]
+  
 a  \, {\rm h_1}  \sin\theta (\partial_\theta Y_{\ell m}) 
\frac{ \ell (\ell+1)  D}{2 AC}
 \, ,
\end{multline}
\begin{multline}
0=
Y_{\ell m}\Bigg[
\psi \left(\frac{B^2 C A'C'}{A}+BC B'C'  +2 B^2 C C'' +\frac{16 B^3 M_2^4 C^2}{\Mpl^2} \right)
\\
+\mathcal{H}_1 \left(
2 \ell (\ell+1)  \left( C B'-\frac{ B\left(C'^2-2 C C''\right)}{C'}   +\frac{16 B^2 M_2^4C^2}{\Mpl^2 C'}  \right) 
+ \frac{2 a m \omega  B}{A}  \left( D C' + D' C \right)\right)
\\
+2 H_0' BCC'
-2 H_0  \ell (\ell+1) C
- H_1 \left(4i\omega \frac{ BC  C'}{A}+  \frac{2 i a m B}{A}  (CD'+C'D) \right)
\\
+ \mathcal{K}' \left(2 C^2 B'+\frac{32 B^2 M_2^4C^3}{\Mpl^2C'}+\frac{4 BC^2 C''}{C'}-2 BCC'\right)
\\
+ \mathcal{K} \left(-\frac{4 \omega ^2 C^2}{A}+2 \left(\ell^2+\ell-2\right) C  -\frac{4 a m \omega  CD}{A} \right)
-  8  a \, {\rm h_0 }  \cos\theta \frac{ \ell (\ell+1) D}{A}
\Bigg]
\\
+  
\frac{2 a B}{A} \sin\theta (\partial_\theta Y_{\ell m})\Bigg[
-i \omega {h_1} \left(DC'+CD'\right)
\\
+{h'_0} \left(DC'+CD'\right)
+ {h_0} \left(C'D'-\frac{2 D A'C'}{A}-\frac{DC'^2}{C} -\frac{2 \ell(\ell+1) D}{B} \right)
\Bigg] \, ,
\end{multline}
\begin{multline}
0=
Y_{\ell m}\Bigg[
\psi  \left(
- \ell (\ell+1) \frac{B \left( C A'+ AC'\right)}{4 A C}+  a m \omega \frac{  B \left(DC'+ CD'\right)}{4 AC}
\right)
\\
+ \mathcal{H}_1 \ell (\ell+1) \left(
    \frac{ \omega ^2 }{2 A}
   -\frac{ \ell (\ell+1)C A'+\left(\ell^2+\ell-2\right) AC'}{2 ACC'}
+ a m \omega   \frac{  DC'+CD'}{2 ACC'} 
\right)
\\
- \frac{1}{2}\ell(\ell+1) H_0'
+ H_0 \left(\ell(\ell+1) \left( \frac{ C'}{4 C} -\frac{ A'}{4 A} \right)
+a m \omega\frac{ D C'- CD'}{4 AC} \right)
+ i \omega H_1\frac{\ell(\ell+1)}{2 A}
\\
 \frac{C }{2 AC'} \mathcal{K}' \left(a m \omega D'-\ell(\ell+1) A'\right)
 + \frac{a m \omega  }{2 A^2} \mathcal{K}  \left(D A'-AD'\right)
 +i a \omega \, {\rm h_1} \cos \theta  \frac{ \ell(\ell+1)   D}{AC}
\\
- 2a \, {\rm h_0'} \cos \theta \frac{ \ell(\ell+1) D}{AC} 
+ a \, {\rm h_0} \cos \theta \, \ell (\ell+1)  \frac{ C \left(D A'-2 AD'\right)+3 ADC'}{A^2 C^2}
\Bigg]
\\
+  
\frac{ a }{2A C} \sin\theta (\partial_\theta Y_{\ell m})\Bigg[ 
i \ell (\ell+1) \omega  D \,  {\rm h_1}
-2 \ell (\ell+1)D \,  {\rm h_0'} 
\\
+ {\rm h_0} \left( \frac{\ell (\ell+1)D A'}{A}+\frac{2 \left(\ell^2+\ell+1\right) DC'}{C}-\left(\ell^2+\ell+2\right) D'
\right)
\Bigg] \, ,
\end{multline}
\begin{multline}
0=
Y_{\ell m}\Bigg[
 \mathcal{K}'' \frac{C^2 A'}{2 AC'}
-  \mathcal{K} \frac{\omega ^2 C}{2 A B}
\\
+\frac{C}{4} \mathcal{K}' \left(
\frac{C'^2 \left(3 B A'+A B'\right)+2 BC''\left(A C'-C A'\right)+2 \left(\ell^2+\ell-4\right) AC'}{A BC'^2}-\frac{4 a m \omega D}{A BC'}-\frac{4 \omega ^2 C}{A BC'}
\right)
\\
+\frac{1}{4} \psi  \left(
\frac{C A' B'+B A'C'+A \left(2 B'C'+2 BC''+\ell^2+\ell-4\right)}{A}-\frac{2 a m \omega D}{A}-\frac{2 \omega ^2C}{A}
\right)
\\
+\frac{1}{4} \psi ' \left(\frac{BC A'}{A}+BC'\right)
-  H_1' \left(\frac{ i a m D}{2A}+\frac{ i \omega C}{A}\right)
\\
-\frac{1}{4} H_1\left(
i \omega\frac{  2 C B'+2  BC'}{A B}
-i a m \frac{   BD A'- AD B'-2  A BD'}{A^2 B}
\right)
\\
\frac{\ell(\ell+1)}{4} \mathcal{H}_1 \Bigg(
-\frac{2 C A' C''}{A C'^2}+\frac{ A'}{A}-\frac{4  \omega ^2 C}{A B C'}+\frac{ B'C'+2 \left(BC''+\ell^2+\ell-4\right)}{BC'}
\\
+ \frac{a m \omega D}{\ell(\ell+1)}  \left(\frac{ A'}{A^2}+\frac{ B' C'-4 \ell (\ell+1)}{A BC'}\right)
\Bigg)
+\frac{1}{4} \mathcal{H}_1'  \left(\frac{2 \ell (\ell+1) C A'}{AC'}+\frac{2 a m \omega  D}{A}\right)
\\
+\frac{1}{2} C H_0'' 
+\frac{1}{4}C H_0' \left(\frac{2  A'}{A}+\frac{ B'}{B}+  \frac{C'}{C}\right)
 -  H_0 \frac{\ell (\ell+1) }{4 B} 
- a \,  {\rm h_0} \cos\theta \frac{ \ell (\ell+1) D}{A BC}
\Bigg]
\\
+  
\frac{ a D}{2A } \sin\theta (\partial_\theta Y_{\ell m})\Bigg[ 
{\rm h_0''}
+ {\rm h_0'} \left(-\frac{3 A'}{2 A}+\frac{B'}{2 B}-\frac{C'}{C}+\frac{2D'}{D}\right)
\\
+ {\rm h_0} \left(\frac{A'C'}{2 AC}-\frac{A'D'}{A D}+\frac{A'^2}{A^2}-\frac{B' C'}{2 BC}+\frac{2}{B C}-\frac{C'D'}{CD}+\frac{C'^2}{C^2}-\frac{C''}{C}\right)
\\
-i \omega  {\rm h_1'}
- i \omega \,  {\rm h_1} \left(\frac{  A'}{2 A}+\frac{  B'}{2 B}\right)
\Bigg] \, ,
\end{multline}
which are obtained from the variation with respect to $H_0$, $H_1$, $H_2$, $\mathcal{H}_1$ and $\mathcal{K}$ respectively, and
\begin{multline}
0=
Y_{\ell m}\Bigg[
-{\rm h_0''} \frac{\ell (\ell+1)}{2 A}
+ {\rm h_0'} \ell (\ell+1) \frac{ B A'-A B'}{4 A^2 B}
+ {\rm h_1'} \frac{i \ell (\ell+1)  }{2 A} \left( \omega + \frac{ a  mD}{ C} \right)
\\
+ {\rm h_0} \frac{\ell (\ell+1) \left(A \left(B'C'+2 \left(BC''+\ell^2+\ell-2\right)\right)-B A'C'\right)}{4 A^2 BC}
\\
+ {\rm h_1} \bigg( 
\frac{i a m \left(-\ell (\ell+1) BCD A'+\ell (\ell+1) ACD B'+2 \left(\ell^2+\ell+2\right) ABCD'-4 A BDC'\right)}{4 A^2 BC^2}
\\
+\frac{i \omega  \left(-\ell (\ell+1) BC A'+\ell (\ell+1) AC B'+2 \ell (\ell+1) A BC'\right)}{4 A^2 BC}
\bigg)
\\
+ a \, \mathcal{K}'' \cos\theta \frac{CD'}{AC'}
+ a \, \mathcal{K}' \cos\theta \left(\frac{D A'}{2 A^2}+\frac{D B'}{2 AB}+\frac{2 \ell(\ell+1) D}{A BC'} -\frac{CC''D'}{AC'^2}+\frac{DC''}{AC'}+\frac{D'}{A} \right)
\\
- a \, \mathcal{K} \cos\theta \frac{\left(\ell^2+\ell+2\right) D}{ABC}
+ \frac{a B}{2A}\psi' \cos\theta \left(\frac{DC'}{C}+D'\right)
+ \frac{a }{2A} H_0' \cos\theta \left(\frac{DC'}{C}-D'\right)
\\
+ \frac{a D}{AC} \psi \cos\theta \left( \ell(\ell+1)  + \frac{B A'C'}{2 A}+B'C'+\frac{C B'D'}{2 D}+BC''\right)
\\
+ a \, \mathcal{H}_1' \cos\theta  \ell (\ell+1)\frac{ CD'-DC'}{ACC'}
+ a \, \mathcal{H}_1 \cos\theta  \frac{\ell (\ell+1)}{A^2 BC^2 C'^2} \Big(
BCD A'C'^2
\\
+A \left(CC' \left(D\left(BC''+2 \ell (\ell+1)\right)-2 BC'D'\right)-BC^2 C''D'+BDC'^3\right) \Big)
\Bigg]
\\
+  
\frac{ a D}{A } \sin\theta (\partial_\theta Y_{\ell m})\Bigg[ 
\mathcal{K}'  \left(\frac{A'}{4 A}+\frac{B'}{4 B}+\frac{\ell(\ell+1)}{BC'}-\frac{CC''D'}{2 DC'^2}+\frac{C''}{2 C'}+\frac{D'}{2 D}\right)
\\
+ \mathcal{K}'' \frac{CD'}{2DC'}
- \mathcal{K} \frac{2 }{BC}
+\psi  \left(\frac{B A'C'}{4 AC}+\frac{B'C'}{2 C}+\frac{B'D'}{4 D}+\frac{BC''}{2 C}+\frac{\ell(\ell+1)}{2 C}\right)
\\
+ \frac{\ell (\ell+1) }{2C^2} \mathcal{H}_1  \left(
\frac{ C A'}{A}+\frac{C \left(BC''+2 \ell (\ell+1)\right)}{BC'}-\frac{\left(\ell^2+\ell+2\right) CD'}{ \ell (\ell+1) D}-\frac{ C^2 C''D'}{DC'^2}+ \frac{2 C'}{\ell (\ell+1)}
\right)
\\
+ \frac{\ell(\ell+1)}{2}\mathcal{H}_1' \left(\frac{ D'}{DC'}-\frac{1}{C}\right)
+ H_0' \left(\frac{C'}{4 C}-\frac{D'}{4D}\right)
\Bigg] \, ,
\end{multline}
\begin{multline}
0=
Y_{\ell m}\Bigg[
\frac{1}{2} {\rm h_1} \left(\frac{2 a \ell (\ell+1) m \omega D}{AC}+\frac{\ell (\ell+1) \omega ^2}{A}  -\frac{\ell(\ell+1)(\ell^2+\ell-2)}{C}\right)
\\
+ \frac{i}{2} {\rm h_0'} \ell (\ell+1)\left(\frac{ m a  D}{AC}+\frac{\omega }{A}\right)
- \frac{i}{2} {\rm h_0} \left(\frac{ a \left(\ell^2+\ell-2\right) m DC'}{AC^2}+\frac{\left(2 a m D'+\ell (\ell+1) \omega  C'\right)}{A C}\right)
\\
- i \omega a \,  \mathcal{K}'  \cos\theta \frac{  CD'}{AC'}
+ \frac{i \omega}{A}  a \,  \mathcal{K}  \cos\theta \left( D'-\frac{  D A'}{A}\right)
-\frac{B}{2A} i\omega a \,  \psi  \cos\theta \left(\frac{  DC'}{C}+   D'\right)
\\
- i \omega a  \,  \mathcal{H}_1 \cos\theta \frac{ \ell (\ell+1)    D'}{AC'}
- a  \, H_1\cos\theta \frac{ \ell (\ell+1) D}{AC}
+\frac{i\omega}{2} a \,   H_0 \cos\theta \left(\frac{  D'}{A}-\frac{   DC'}{AC}\right)
\Bigg]
\\
+  
\frac{ a D}{4AC } \sin\theta (\partial_\theta Y_{\ell m})\Bigg[ 
- 2 i \omega \mathcal{K}' \frac{ C^2D' }{DC'}
+2i \omega  \mathcal{K} C \left(\frac{   D'}{D}-\frac{ A'}{A}\right)
- 2\ell(\ell+1) H_1 
\\
- i \omega  \psi  B \left( C'+\frac{CD'}{D}\right)
- 2 i \omega \ell (\ell+1) \mathcal{H}_1\frac{ CD'}{DC'}  
+i \omega H_0 \left(\frac{ CD'}{D}- C'\right)
\Bigg] \, ,
\end{multline}
which follow instead from the variation with respect to ${\rm h_0}$ and ${\rm h_1}$, respectively.
It is clear from the previous equations that, as opposed to the non-rotating case \cite{Franciolini:2018uyq,Franciolini:2018aad}, the even and odd equations are now coupled. The dependence on the spherical harmonics can be removed by acting on each  equation with  $\int\D \Omega \, Y_{\ell'm'}^*(\theta,\phi)$ and using the  identities \eqref{eq:recurrence1a} and \eqref{eq:recurrence1b}.
It becomes then clear that the even-odd mixing terms in the previous equations are between modes of different parity whose azimuthal quantum numbers  differ by $1$. In other words, at linear order in $a$, the coupling is between an even $\ell$-mode and an odd $(\ell\pm1)$-mode, and vice versa \cite{1991RSPSA.433..423C}.
It is then easy to find the combinations of the previous equations that are algebraic in the fields $H_0$, $H_1$, $\mathcal{H}_0$ and ${\rm h}_0$. Solving for these variables and plugging the solutions into the remaining independent equations, one finds the final set of coupled equations for the three propagating fields $\psi$, $\mathcal{K}$ and ${\rm h}_1$.
Dropping the even-odd mixing, the equation for ${\rm h}_1$ decouples and reduces to \eqref{oddeftgkerr} in the main text.
Note also that the same manipulations have been performed to obtain the systems of equations  \eqref{cgeqpsieven} and \eqref{cgeqpsieven-2}, with the huge simplification that, in the context of \eqref{cgeqpsieven} and \eqref{cgeqpsieven-2}, $a=0$ and there is no mixing between even and odd fields.

\paragraph{Modes with $\ell=1$.} When $\ell=1$, only the scalar mode propagates, although its equation of motion can still couple to the odd $\ell=2$ metric fluctuations. To derive the linearized   equation, one can proceed as before with the simplification now    that, when  $\ell=1$,  $\mathcal{K}$ is redundant in the parametrization \eqref{even metric perturbations version 0} and can be set to zero. Similar field redefinitions  (see eq.~\eqref{H2fr}) and manipulations to the ones above allow  to integrate out $H_0$, $H_1$ and $\mathcal{H}_0$, and find the equation of motion for the dynamical  field $\psi$ (see also Refs.~\cite{Franciolini:2018uyq,Franciolini:2018aad}).

\paragraph{Mode with $\ell=0$.} The monopole $\ell=0$ can be derived in complete analogy. Note that, since it describes a spherically symmetric perturbation, its equation of motion will not contain any linear term proportional to the spin parameter $a$. The field equation for $\ell=0$ will be therefore identical to the one derived in the non-rotating case in Refs.~\cite{Franciolini:2018uyq,Franciolini:2018aad}.

\bibliographystyle{JHEP}
\addcontentsline{toc}{section}{References}
\bibliography{biblio}

\end{document}